%% file: complexity.tex
\documentclass[a4paper,marginparwidth=0pt,twocolumn]{article}
\pdfoutput=1
\usepackage{graphicx}
\usepackage{color}
\usepackage[top=3.2cm, bottom=3cm, left=1.5cm, right=1.5cm]{geometry}
\usepackage{amsmath}
\usepackage{amssymb}
\usepackage{authblk}
\usepackage{float}

\usepackage{fancyhdr}
\usepackage{titlesec}
\titleformat*{\section}{\large\bfseries}
\titleformat*{\subsection}{\normalsize\bfseries}
\titleformat*{\subsubsection}{\normalsize}

\newcommand\abs[1]{\left|#1\right|}

\usepackage[normalem]{ulem}

\input{symbollines.tex}

\begin{document}
\title{\LARGE \bfseries Complexity of localised coherent structures in a boundary-layer flow}

\author[1,2]{\small{Taras Khapko}}
\author[3]{Yohann Duguet}
\author[4,5]{Tobias Kreilos}
\author[1,2]{Philipp Schlatter}
\author[4,6]{Bruno Eckhardt}
\author[1,2]{Dan S. Henningson}

\affil[1]{Linn\'e FLOW Centre, KTH Mechanics, Osquars Backe 18, SE-100 44 Stockholm, Sweden}
\affil[2]{Swedish e-Science Research Centre (SeRC)}
\affil[3]{LIMSI-CNRS, UPR 3251, F-91403 Orsay Cedex, France}
\affil[4]{Fachbereich Physik, Philipps-Universit\"at Marburg, Renthof 6, D-35032 Marburg, Germany}
\affil[5]{Max Planck Institute for Dynamics and Self-Organization, Am Fassberg 17, D-37077 G\"ottingen, Germany}
\affil[6]{J.M. Burgerscentrum, Delft University of Technology, Mekelweg 2, NL-2628 CD Delft, The Netherlands}

\twocolumn[
  \begin{@twocolumnfalse}

\maketitle

\abstract{We study numerically transitional coherent structures in a boundary-layer flow with homogeneous suction at the wall
(the so-called asymptotic suction boundary layer ASBL). The dynamics restricted to the laminar--turbulent separatrix is investigated in a
spanwise-extended domain that allows for robust localisation of all edge states. We work at fixed Reynolds number and study the edge states as a
function
of the streamwise period. We demonstrate the complex spatio--temporal dynamics of these localised states, which exhibits
multistability and undergoes complex bifurcations leading from periodic to chaotic regimes. It is argued that in all regimes the
dynamics restricted to the edge is essentially low-dimensional and non-extensive. \vspace{1cm}
}
  \end{@twocolumnfalse}
]

\section{Introduction}
\label{sec:intro}

\begin{figure}
\centering
\includegraphics[scale=0.4]{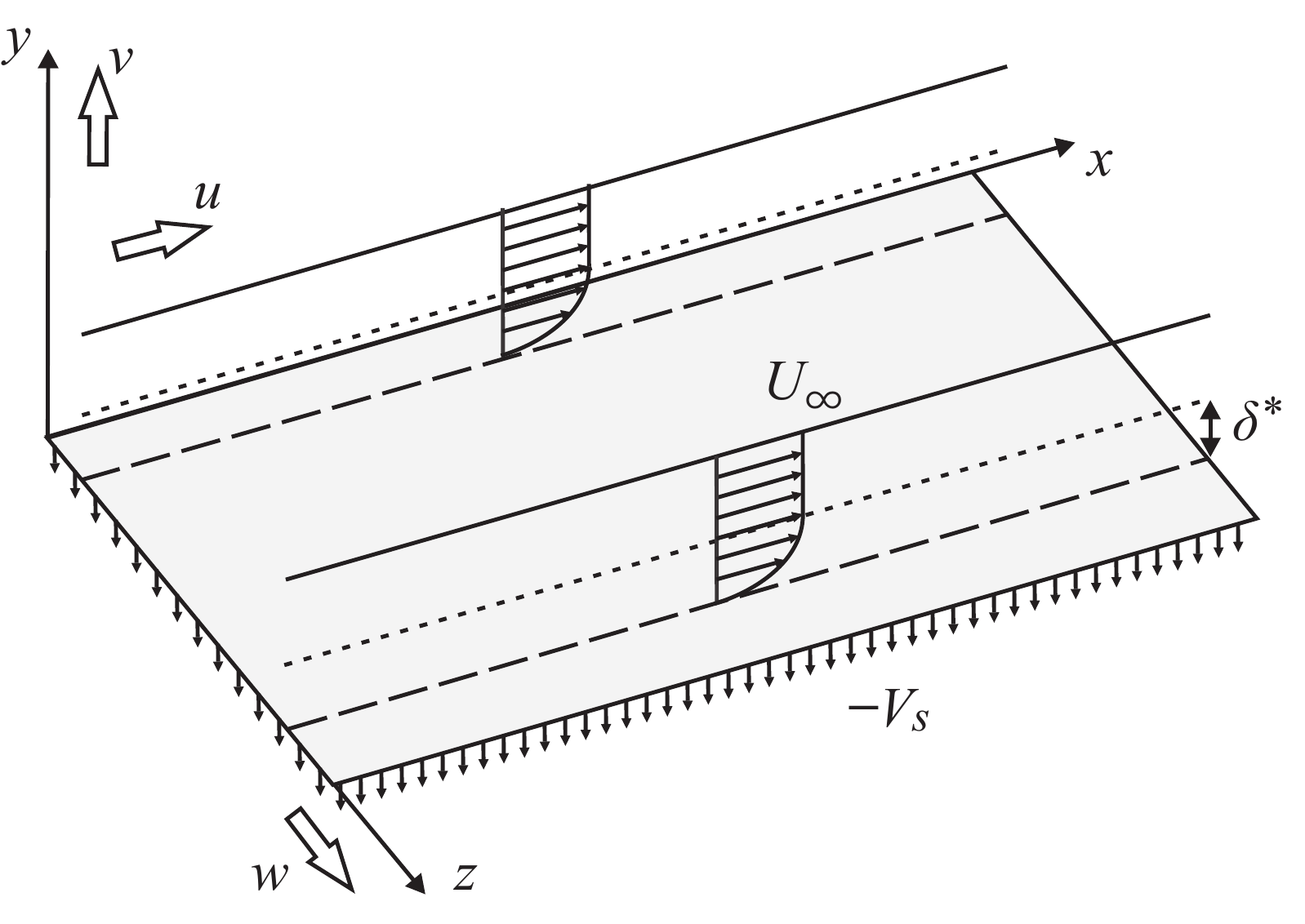}
\caption{\label{fig:asbl} \small Sketch of the asymptotic suction boundary layer. }
\end{figure}

\begin{figure}
\centering
\includegraphics[scale=0.4]{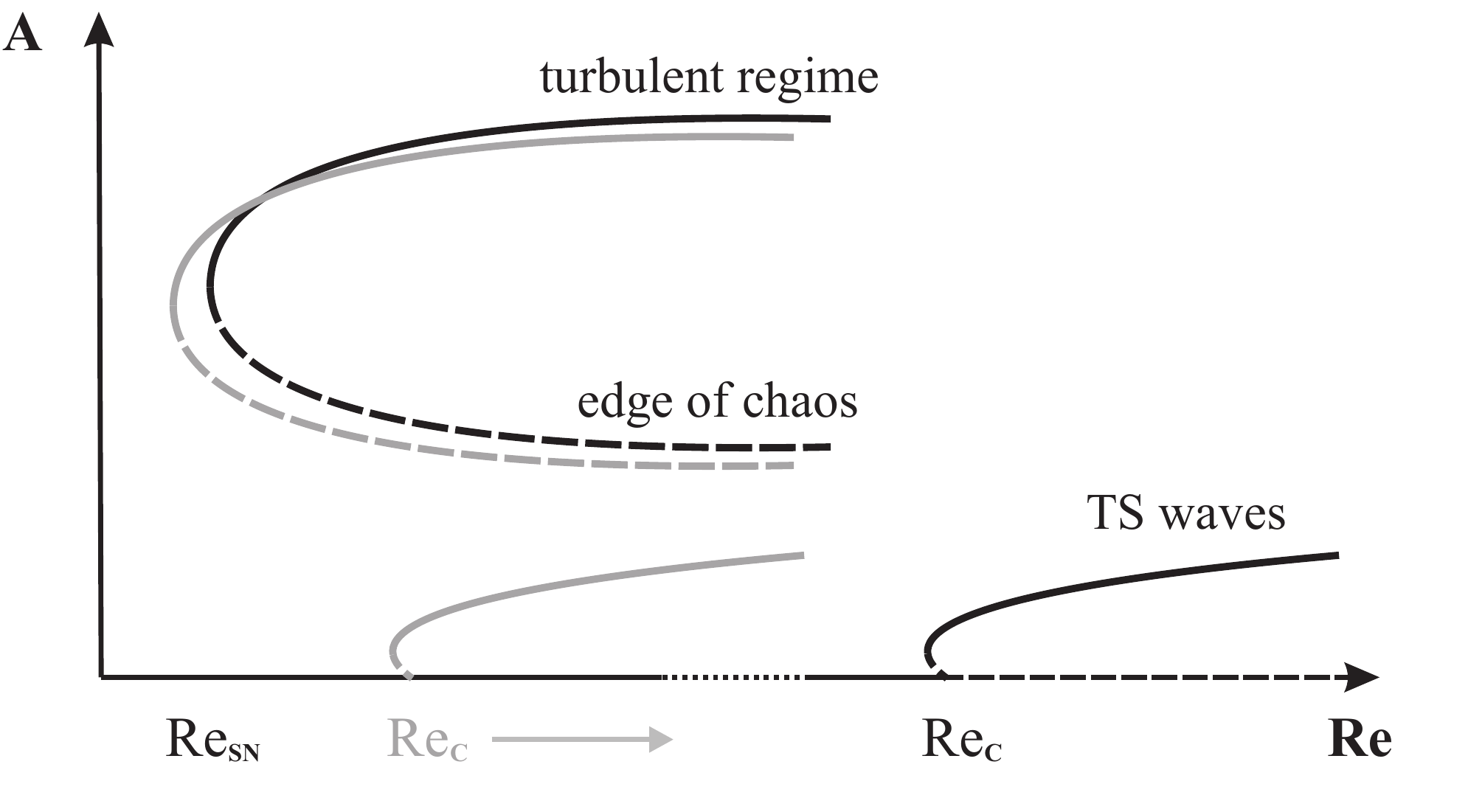} \vspace{.16cm}
\caption{\label{fig:bl_bif} \small Schematic bifurcation diagram for the uncontrolled (grey) and controlled (black) boundary layer.}
\end{figure}

Understanding how boundary-layer flows become turbulent is of crucial importance for many applications, in particular for all aeronautic
purposes, because
of the increased drag associated with turbulent fluctuations. Since in most applications drag needs to be kept as low as possible, control strategies
aim at delaying the transition from laminar to turbulent. The problem of transition to turbulence has long been addressed using linear stability
theory, where typically one looks for a critical value of a governing parameter above which the laminar base flow loses its stability with respect to
infinitesimal disturbances. For instance, in the case of the incompressible flat plate Blasius boundary-layer flow, a Reynolds number
$Re=U_{\infty}\delta / \nu$ can be constructed using the free-stream velocity $U_{\infty}$, the local displacement thickness $\delta$ at the given
streamwise location, and the kinematic viscosity $\nu$ of the fluid. The critical value of $Re$ given by linear stability theory is $Re_c \approx
520$, which corresponds to an exponential amplification of Tollmien--Schlichting (TS) waves \cite{schmid_henningson_2001}. TS waves are ``connected"
to
the base flow, meaning that there is a continuous path in parameter space linking them to the laminar state. TS waves can actually be triggered at
lower (subcritical) $Re$, resulting in a
finite-amplitude instability of the base flow to such waves. The
``classical" transition route, typical of weakly disturbed environments, corresponds to the sequence of secondary bifurcations undergone by these
waves \cite{herbert_1988}. Still according to linear stability theory, a control strategy that would shift the linear stability threshold would
be considered efficient. We consider here the case of an
incompressible boundary-layer flow stabilised by homogeneous suction at the wall. Suction
counteracts the spatial development of the boundary layer, leading asymptotically to a steady laminar base flow independent of the planar coordinates,
called Asymptotic Suction Boundary-Layer flow (ASBL) \cite{schlichting_1987}, see fig.~\ref{fig:asbl}. Since suction shifts the linear stability
threshold to $54,370$ \cite{hocking_1975} it would qualify as an efficient control. However, in this flow as well as in other shear flows, a
second, nonlinear path to turbulence, so-called bypass transition, can interfere with this control strategy.
In the presence of stronger background fluctuations (noise, incoming turbulence, localised forcing, \emph{etc.}), a drastically different, fully
nonlinear picture emerges for transition, termed ``bypass route to turbulence". Schematically, additional solutions of the governing equations that
are
\emph{not} connected to the base flow appear at a finite value of $Re=Re_{SN}$ through a saddle--node bifurcation
\cite{dauchot_manneville_1997,manneville_2004}. The upper branch corresponds to solutions with larger drag, representative of a stable turbulent
regime. Its counterpart, the lower branch, corresponds to an unstable separatrix dividing the phase space into two basins of attraction for the laminar
and turbulent state, respectively. The value of $Re_{SN}$ cannot be predicted by linear stability theory, and generally  $Re_{SN}<Re_c$. Perhaps more
importantly, control strategies based on linear theory also fail at predicting whether $Re_{SN}$ is affected at all by the control, see
fig.~\ref{fig:bl_bif}. There is recent evidence that $Re_{SN}$ for both Blasius and ASBL flows is as low as $300$ \cite{levin_henningson_2007}. This
points out the need for a nonlinear description of transition processes extended to controlled boundary-layer flows.

The existence of a lower branch not connected to the base flow is typical of a wider class of transition scenarios classified as ``subcritical
transition", common to many wall bounded flows such as circular pipe flow, square duct flow, plane Couette flow, plane channel flow, see for instance
\cite{manneville_2004,eckhardt_schneider_hof_westerweel_2007}. As mentioned before, the phase space is split into two regions, one where trajectories
immediately return to the laminar fixed point, the other where the turbulent state is reached. Detailed investigations of the dynamics along the
associated separatrix have become popular in the last decade owing to the concept of ``edge state" and ``edge manifold"
\cite{skufca_yorke_eckhardt_2006}. The edge manifold (or simply ``edge") is not only the geometrical separatrix in phase space, but also an invariant
subspace for the flow. Within this invariant manifold there are attractors, so-called edge states. In the simplest case, the edge state is unique and
corresponds to a simple unstable solution of the system, such as a fixed point or a limit cycle, and its stable manifold (the edge) separates two
attractors in state space: the laminar state and the turbulent state. For initial conditions close enough to the edge, trajectories will then approach
the neighbourhood of the edge state and later leave towards one of the two attractors depending on which ``side" of the edge the initial condition
lies. An efficient and intrinsically nonlinear control strategy should then consist of targeting the edge state in order to shift the dynamics from
the turbulent side of the edge to the laminar one \cite{kawahara_2005}.

When the system is not geometrically restricted by periodic boundary conditions, the edge state corresponds to a coherent structure of the flow
that is always spatially localised
\cite{mellibovsky_meseguer_schneider_eckhardt_2009,duguet_schlatter_henningson_2009,schneider_marinc_eckhardt_2010,duguet_willis_kerswell_2010,duguet_schlatter_henningson_eckhardt_2012}.
Families of localised structures in plane Couette flow have been connected to the snaking bifurcation scenarios
\cite{knobloch_2008,schneider_gibson_burke_2010,duguet_maitre_schlatter_2011}.

Several difficulties arise, making this picture often incomplete. The first one, typical of low values of $Re$, occurs when the turbulent state is a
chaotic saddle rather than an attractor. The stable manifold of the edge state is in that case entangled with the turbulent dynamics in a complex way,
a manifestation being a finite probability of relaminarising for almost any turbulent trajectory. Such behaviour is in general linked to a boundary
crisis occurring at some value of $Re$ slightly above $Re_{SN}$
\cite{mellibovsky_eckhardt_2012,kreilos_eckhardt_2012,avila_mellibovsky_rolland_hof_2013} involving the stable manifold of the edge state. At higher
$Re$, sudden relaminarization events are no longer observed in practice, either because turbulent lifetimes are too long on average, or because
spatial proliferation of turbulent fluctuations makes the local probability for relaminarization only weakly relevant
\cite{manneville_2009,avila_moxey_lozar_avila_barkley_hof_2011}. Another difficulty, which we will address here, is related to simple general
questions such as: ``what is the expected nature of the edge state?", ``can it be chaotic?", ``is the edge state unique?"  and ``is it robust to
changes in the parameters?". Some of these questions have been addressed using a simple low-dimensional phenomenological model in
ref.~\cite{vollmer_schneider_eckhardt_2009}. The concept of edge state as a relative attractor is an asymptotic concept only. As edge-tracking
algorithms are iterative by nature, numerical evidence for chaotic edge states is hard to justify since any erratic edge trajectory is always
potentially a transient approach to a more simple invariant state. Chaotic edge states have been nevertheless reported several times in the literature
\cite{schneider_eckhardt_yorke_2007,duguet_schlatter_henningson_2009,schneider_marinc_eckhardt_2010} and are likely to be typical if not generic in
extended domains. Phase-space coexistence of different (\emph{i.e.}\ not symmetry-related) edge states has been reported too
\cite{duguet_willis_kerswell_2008,khapko_kreilos_schlatter_duguet_eckhardt_henningson_2013}. These questions are important from a fundamental point of
view because the separatrix forms part of the skeleton of the phase space. They are also crucial from a practical point of view, a multitude of edge
states calling for a non-trivial generalisation of the nonlinear control strategy suggested in ref.~\cite{kawahara_2005}. Apart from transition
processes, studying coherent structures on the edge can also teach us about the dynamics of near-wall turbulence
\cite{jimenez_kawahara_simens_Nagata_shiba_2005}.

Recent numerical studies have focused on edge states in various models of boundary-layer flows
\cite{cherubini_depalma_robinet_bottaro_2011,duguet_schlatter_henningson_eckhardt_2012,biau_2012}, in particular in the case of ASBL
\cite{kreilos_veble_schneider_eckhardt_2013,khapko_kreilos_schlatter_duguet_eckhardt_henningson_2013}. We here continue to explore the rich
dynamics of edge states in the ASBL as a prototype for controlled
boundary-layer flows. Following the strategy adopted in Couette flow \cite{schneider_marinc_eckhardt_2010}, as the next step after
studying small
boxes \cite{kreilos_veble_schneider_eckhardt_2013} and prior to considering large ones, we perform numerical simulations in a
periodic domain that is wide enough to display
spatial localisation in the spanwise direction. Let us denote by $L_x$ the streamwise length of the domain, which in case of
periodic boundary conditions defines the fundamental streamwise wavenumber in a given computational domain. Unlike more classical
procedures where the governing parameter varied is $Re$, here we keep the value of $Re$ constant and consider the streamwise wavelength $L_x$ as a
varying parameter. Various types of dynamics on the edge emerge from this numerical exploration, among them period doubling, multistability and
Pomeau--Manneville intermittency that are surprisingly reminiscent of many simpler systems such as coupled logistic maps
\cite{kuznetsov_kuznetsov_saraev_1993}.

The paper is structured as follows. In sect.~2 we introduce the essential features of the ASBL and recall the numerical methods combined together to
identify edge states. Sect.~3 begins with examples of the low-$Re$ dynamics of the flow and the evidence for a chaotic saddle related to the stable
manifold of the edge state. The various states identified in Khapko \emph{et al.}\ $2013$
\cite{khapko_kreilos_schlatter_duguet_eckhardt_henningson_2013} are then tracked in parameter space \emph{vs.}\ changes in $L_x$. Sect.~4 is devoted
to a discussion of the results and suggests a more general picture of the dynamics on the basin separatrix. Finally the relevance of such localised
states for the study of the turbulent regime will be discussed.

\section{Flow case and numerical methods}
\label{sec:numerics}

The asymptotic suction boundary layer (ASBL) is a zero-pres\-sure-gradient boundary-layer flow above a flat plate at which constant
homogeneous suction is applied. The flow is governed by the incompressible Navier--Stokes equations
\begin{equation}
\frac{\partial \tilde{\mathbf{u}}}{\partial \tilde{t}} + (\tilde{\mathbf{u}} \cdot \tilde{\mathbf{\nabla}})\tilde{\mathbf{u}} = -\frac 1
{\rho}
\tilde{\mathbf{\nabla}}\tilde{p} + \nu
\tilde{\mathbf{\nabla}}^2\tilde{\mathbf{u}} \ ,
\end{equation}
together with the continuity equation
\begin{equation}
\tilde{\mathbf{\nabla}} \cdot \tilde{\mathbf{u}} = 0 \ .
\end{equation}
Here $\tilde{\mathbf{u}}=(\tilde{u},\tilde{v},\tilde{w})$ is the velocity field of the flow in the streamwise $\tilde{x}$, wall-normal $\tilde{y}$ and
spanwise $\tilde{z}$ directions, respectively, $\tilde{p}$
stands for the pressure, $\rho$ for the fluid density and $\nu$ for the kinematic viscosity. The corresponding boundary conditions are
\begin{subequations}
\begin{align}
{(\tilde{u},\tilde{v},\tilde{w})}_{\tilde{y}=0} & = (0,-V_S,0) \ , \\
{(\tilde{u},\tilde{v},\tilde{w})}_{\tilde{y}=\infty} & = (U_\infty,-V_S,0) \ ,
\end{align}
\label{BCs}
\end{subequations}
\hspace{-0.15cm}where $U_\infty$ and $V_S$ are the fixed free-stream and suction velocities.

The system admits a steady laminar solution with constant boundary-layer thickness,
\begin{subequations}
\begin{align}
(\tilde{U},\tilde{V},\tilde{W}) &= (U_\infty(1-e^{-\tilde{y}V_S/\nu}), -V_S, 0) \ , \\
\tilde{p} &= const \ .
\end{align}
\end{subequations}

The Reynolds number $Re=U_\infty \delta^* / \nu$ is based on the laminar displacement thickness
\begin{equation}
\delta^*=\int_0^\infty \! \left(1-\tilde{u}(\tilde{y})/U_\infty\right) \, \mathrm{d} \tilde{y} \ ,
\end{equation}
which in this case is given analytically by $\delta^*=\nu/V_S$. Accordingly, the
Reynolds number is given by the ratio of the free-stream velocity and the suction velocity, $Re=U_\infty / V_S.$ The free-stream
velocity $U_\infty$ and the boundary-layer thickness $\delta^*$ are used as characteristic units for the non-dimensionalisation, and we
write non-dimensional quantities without tilde.

For the current study direct numerical simulations of ASBL are performed using a fully spectral code in a channel geometry
\cite{chevalier_schlatter_lundbladh_henningson_2007} of finite wall-normal extent $[0,L_y]$. The details of the numerical method are discussed
in
ref.~\cite{khapko_kreilos_schlatter_duguet_eckhardt_henningson_2013}. Whereas the streamwise extent $L_x$ of the
computational domain is varied between $4 \pi$ and $6 \pi$ (in units of $\delta^{*}$), the height $L_y$ and the width $L_z$
are held constant, as well as the Reynolds number which is set to $Re=500$ except when indicated. As shown in
ref.~\cite{khapko_kreilos_schlatter_duguet_eckhardt_henningson_2013}, $L_y=15$ and $L_z=50$ are sufficient to accurately catch the localisation
properties of the edge states. A resolution of $N_x\times N_y \times N_z = 48 \times 129 \times 192$ spectral modes was found suitable for the
investigation of edge states within this numerical domain, resulting in a system with $N = 4N_xN_yN_z \approx 4.8 \times 10^6$ degrees
of freedom for the velocities and pressure.

In order to track the dynamics on the laminar--turbu\-lent separatrix, a bisection is performed along the line joining an arbitrary initial condition
to the laminar state \cite{skufca_yorke_eckhardt_2006}. Each individual trajectory is followed until it approaches either the turbulent or the laminar
one. Such approaches are detected using predetermined thresholds for the quantity $v_{\mathrm{rms}}$, which represents root-mean-square of the
wall-normal velocity fluctuations. This iterative procedure results in a trajectory that shadows the laminar--turbulent boundary for arbitrary long
times. By iterating for a sufficiently long time we can determine the nature of the relative attractor on the edge, \emph{i.e.}\  the edge state.

\section{Results}
\label{sec:results}

\subsection{Low-$Re$ transient turbulence}

\begin{figure}
\centering
\includegraphics[width=.9\linewidth]{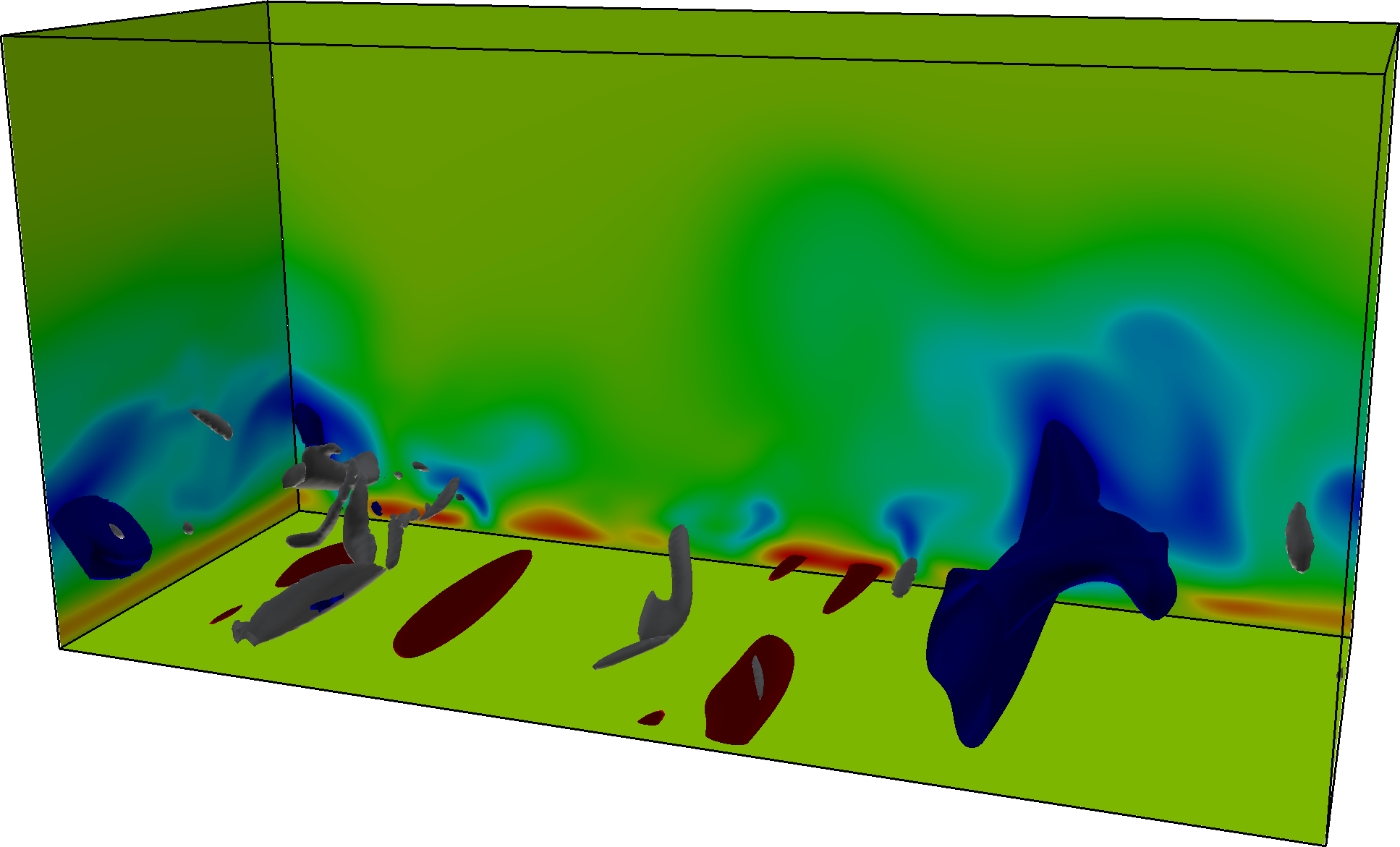}
\caption{\label{fig:turbulent_asbl} \small Three-dimensional visualisation of turbulent ASBL for $Re=270$ in a numerical domain with $L_x=6\pi$. Blue
(low-speed streak) and red (high-speed streak) are the isosurfaces of streamwise velocity fluctuations $u'=-0.4$ and $u'=0.2$. Vortices are visualised
using the $\lambda_2$ criterion \cite{jeong_hussain_1995} with the isosurface $\lambda_2=-0.01$. Flow from lower left to upper right.}
\end{figure}

We begin with a short description of the turbulent state obtained by numerical simulation. Our investigations so far show that transition to
turbulence can be observed in the ASBL at least for Re $\gtrsim 250$ provided that the computational domain is large enough.
Fig.~\ref{fig:turbulent_asbl} shows a typical turbulent state at $Re=270$ and for $L_x=6\pi$. A slightly higher computational domain with $L_y=25$ was used in order to account for
the thickening of the boundary layer typical for the turbulent ASBL \cite{schlatter_orlu_2011}. The flow features clear coherent structures close to the
wall in the form of high-speed streaks (shown in red) very near the lower wall, and low-speed streaks (blue) further up into the flow. The flow shows
however no sign of spanwise localisation. The redistribution of momentum in the vertical direction is mainly due to the advection by
streamwise vortices.

\begin{figure}
\centering
\includegraphics[width=.8\linewidth]{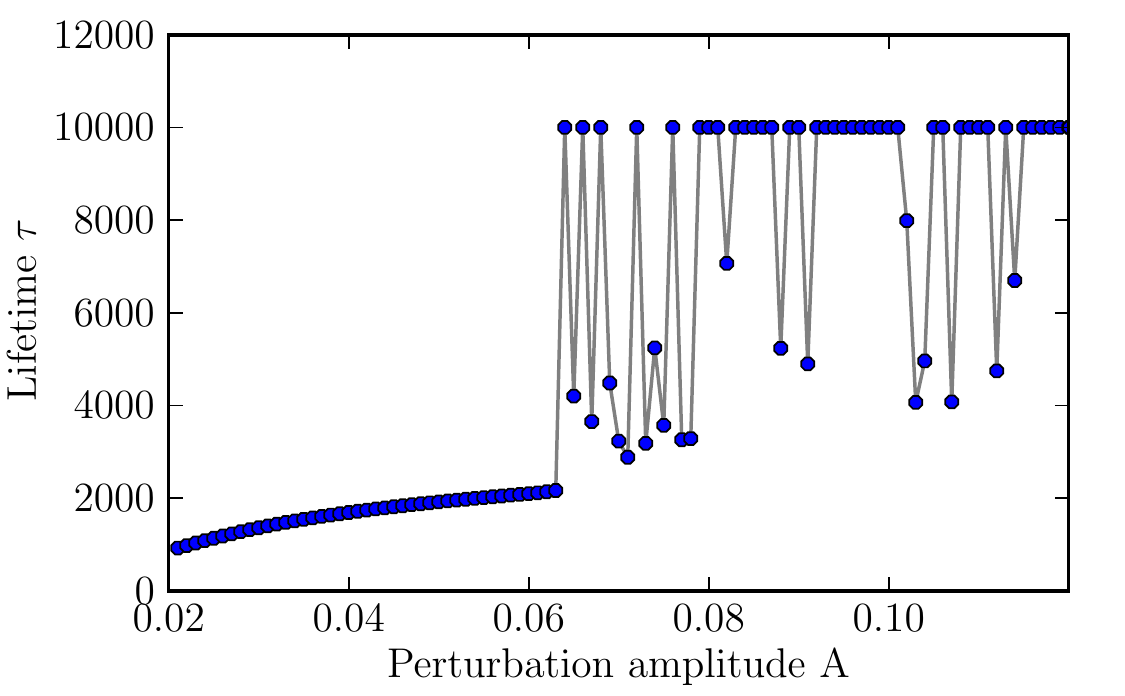}
\caption{\label{fig:lifetimes_asbl} \small Lifetimes associated to trajectories starting from the initial condition $\mathbf{u}_{A}$, \emph{vs.}\ the
disturbance
amplitude $A$, at $Re=270$. $A=1$ corresponds to the snapshot shown in fig.~\ref{fig:turbulent_asbl}. The maximum observation time is here $10,000$
time units.}
\end{figure}

The turbulent regime in the chosen computational domain at such low values of $Re$ is actually only metastable and the flow can rapidly relaminarise
after a (sometimes extremely long) finite time. An investigation of lifetimes was performed at $Re=270$ by considering the perturbation $\mathbf{u'}$
to the laminar state $\mathbf{U}$ displayed in fig.~\ref{fig:turbulent_asbl}, rescaling its amplitude by a factor $A$, considering new initial
conditions
$\mathbf{u}_{A}(t=0)=\mathbf{U} + A\mathbf{u'}$ and finally measuring the lifetime of the turbulent regime as a function of $A$. Below a given
threshold in $A$, all trajectories decay rapidly to the laminar state. Above this threshold $A \approx 0.06$ lifetimes increase dramatically and show
huge fluctuations, with isolated initial conditions persisting forever. However, averaged over smooth sets of initial conditions, the mean lifetimes
remain finite, indicating that the edge has been crossed \cite{schneider_eckhardt_yorke_2007}. In fig.~\ref{fig:lifetimes_asbl} the lifetimes are
displayed as a function of $A$ showing a
very fractal landscape. This situation, analogous to most subcritical
shear flows at low enough values of $Re$, suggests the existence of a chaotic saddle
\cite{schmiegel_eckhardt_1997}. As recently
shown for the case of pipe flow and plane Couette flow \cite{mellibovsky_eckhardt_2012,kreilos_eckhardt_2012,avila_mellibovsky_rolland_hof_2013}, the
creation of this chaotic saddle is due to a boundary crisis as $Re$ is increased. This example of global bifurcation is typical in dynamical systems
\cite{alligood_sauer_yorke_1996} and emerges when an already existing chaotic attractor collides with its basin boundary, which
is the stable manifold of the edge state. In all studies of shear flows, the chaotic attractor preceding the boundary crisis emerges from a sequence
of local bifurcations from an ``upper branch" state originating from a saddle--node bifurcation. The other (unstable) state originating from that
saddle--node is precisely the edge state. In practice, the studies in refs.~\cite{kreilos_eckhardt_2012,avila_mellibovsky_rolland_hof_2013} all
started with the identification of the edge state on the basin boundary, followed using continuation techniques down to the saddle--node bifurcation,
where both lower and upper branches are created. This points out the importance of the edge state as the backbone of the turbulent dynamics, not only
of the basin boundary. For larger values of $Re$, lifetimes tend to increase on average. Whether the turbulent regime remains transient or not at
larger values of $Re$ is beyond the scope of this study, however the concept of edge state remains robust. Note that the numerical continuation
performed in former studies demands for technical reasons an edge state with trivial time dependence. In the remainder of the paper we investigate
edge states in spanwise-extended ASBL and their dynamics.

\subsection{Periodic edge states for $L_x=6\pi$}

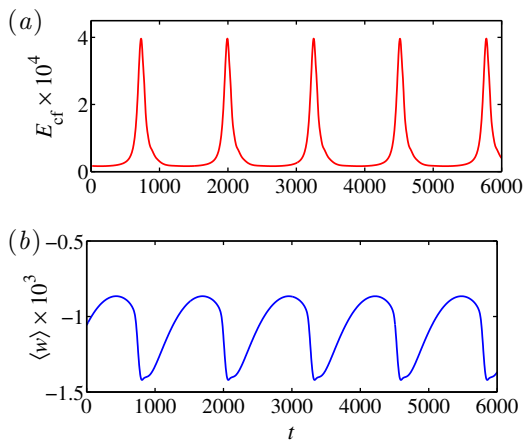
\begin{figure}
\centering
\includegraphics[scale=0.33]{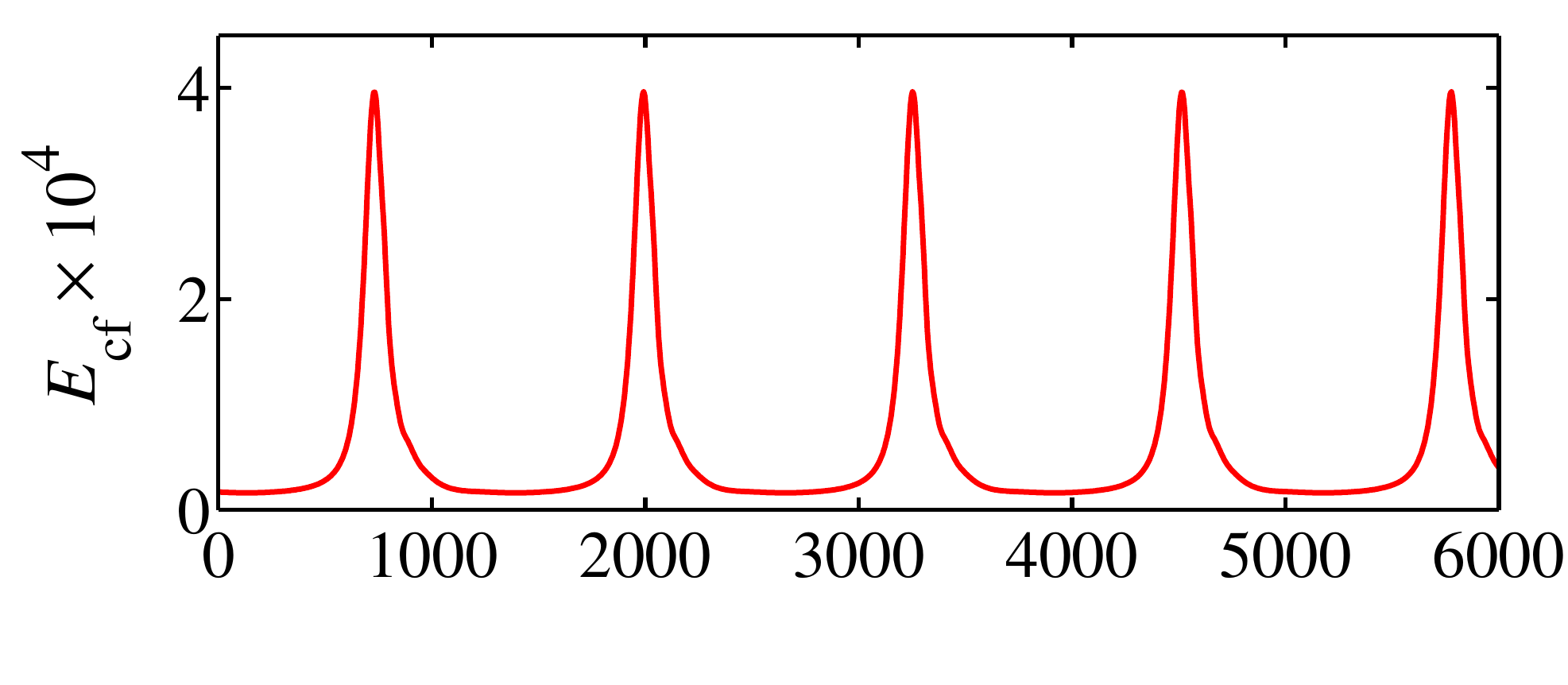}
\includegraphics[scale=0.33]{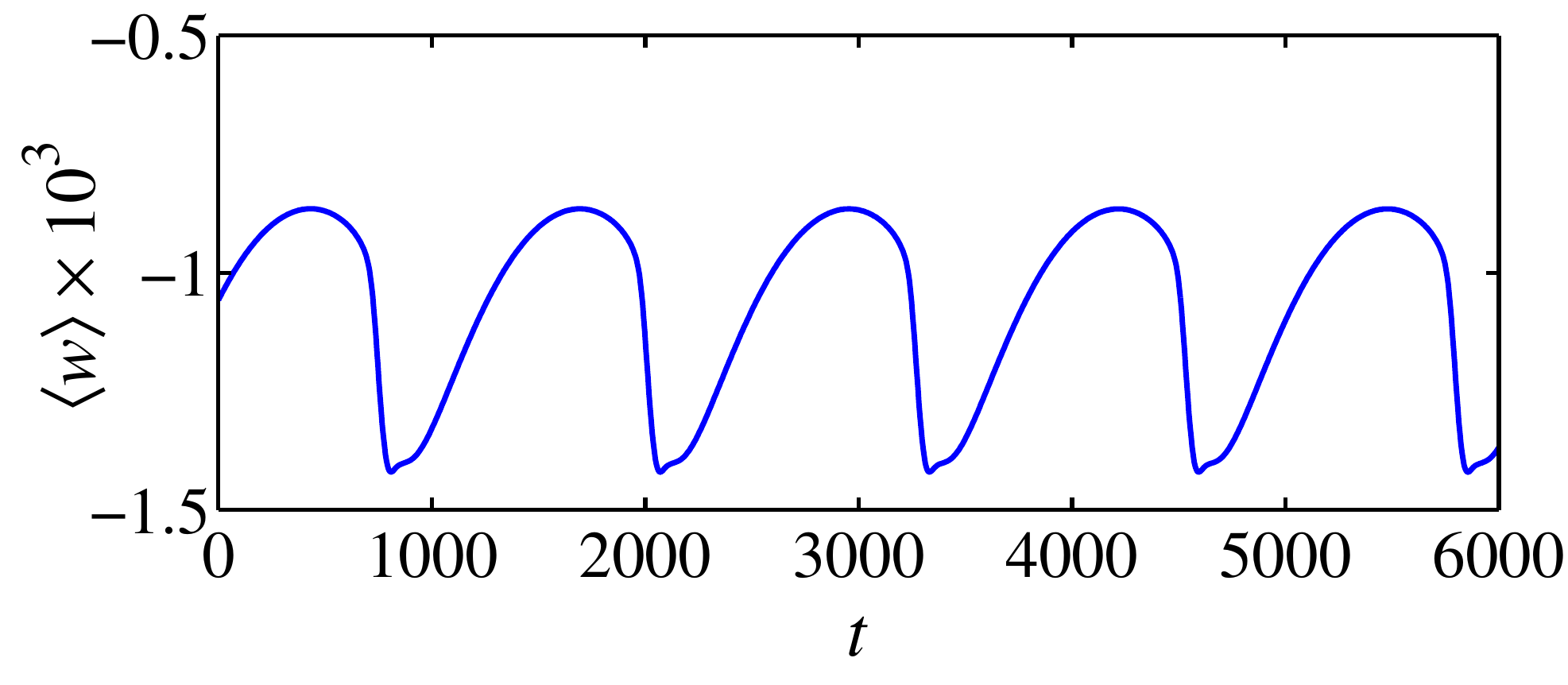}
\begin{picture}(0,0)
\put(-195,158){(\textit{a})}
\put(-195,75){(\textit{b})}
\end{picture}
\caption{\label{fig:6pi_l_Ecf} \small Time series for the left-shifting state (L) at $L_x=6\pi$: (\textit{a}) cross-flow energy $E_{\mathrm{cf}}$;
(\textit{b}) mean spanwise velocity $\langle w \rangle$.}
\end{figure}

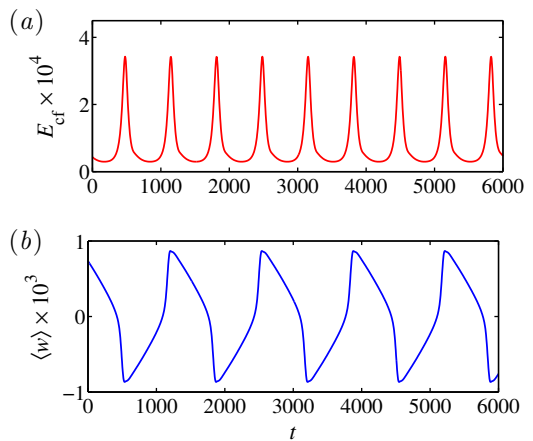
\begin{figure}
\centering
\includegraphics[scale=0.33]{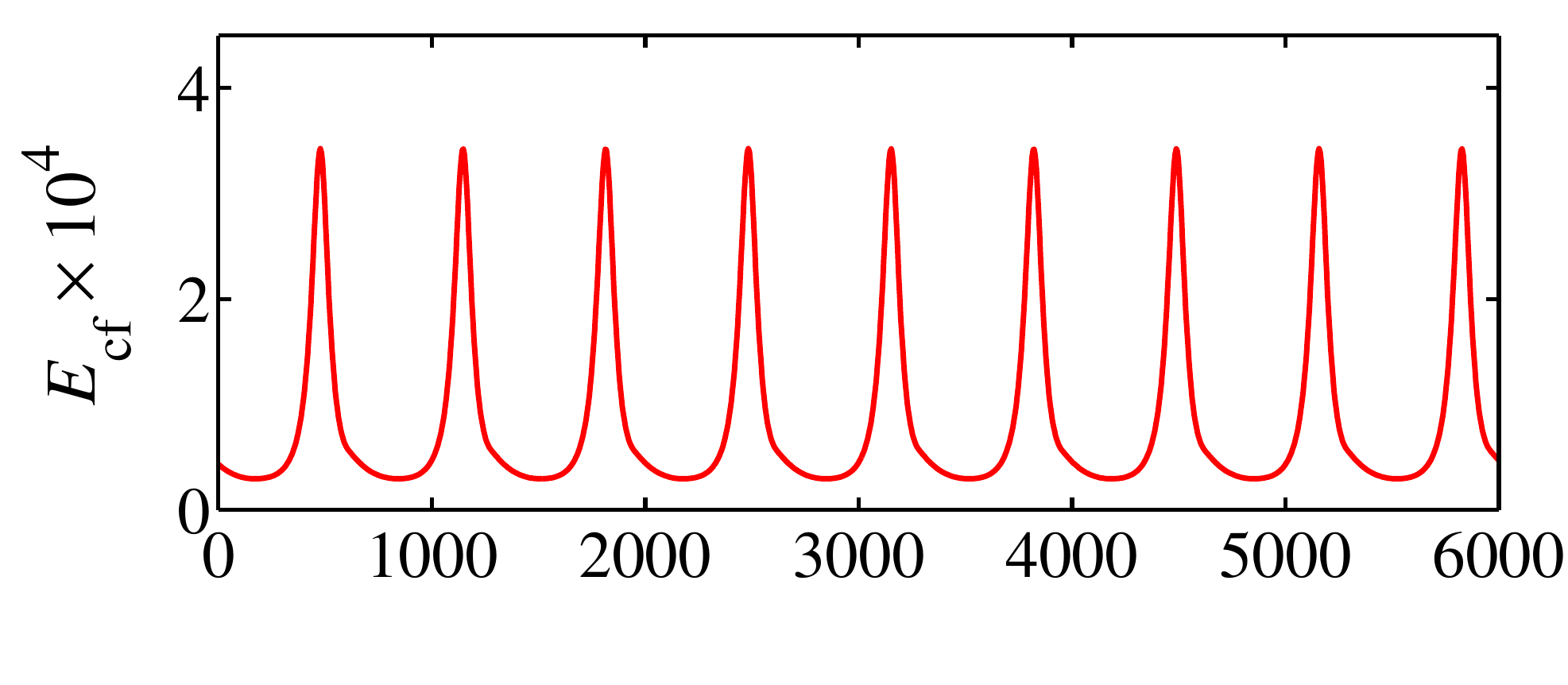}
\includegraphics[scale=0.33]{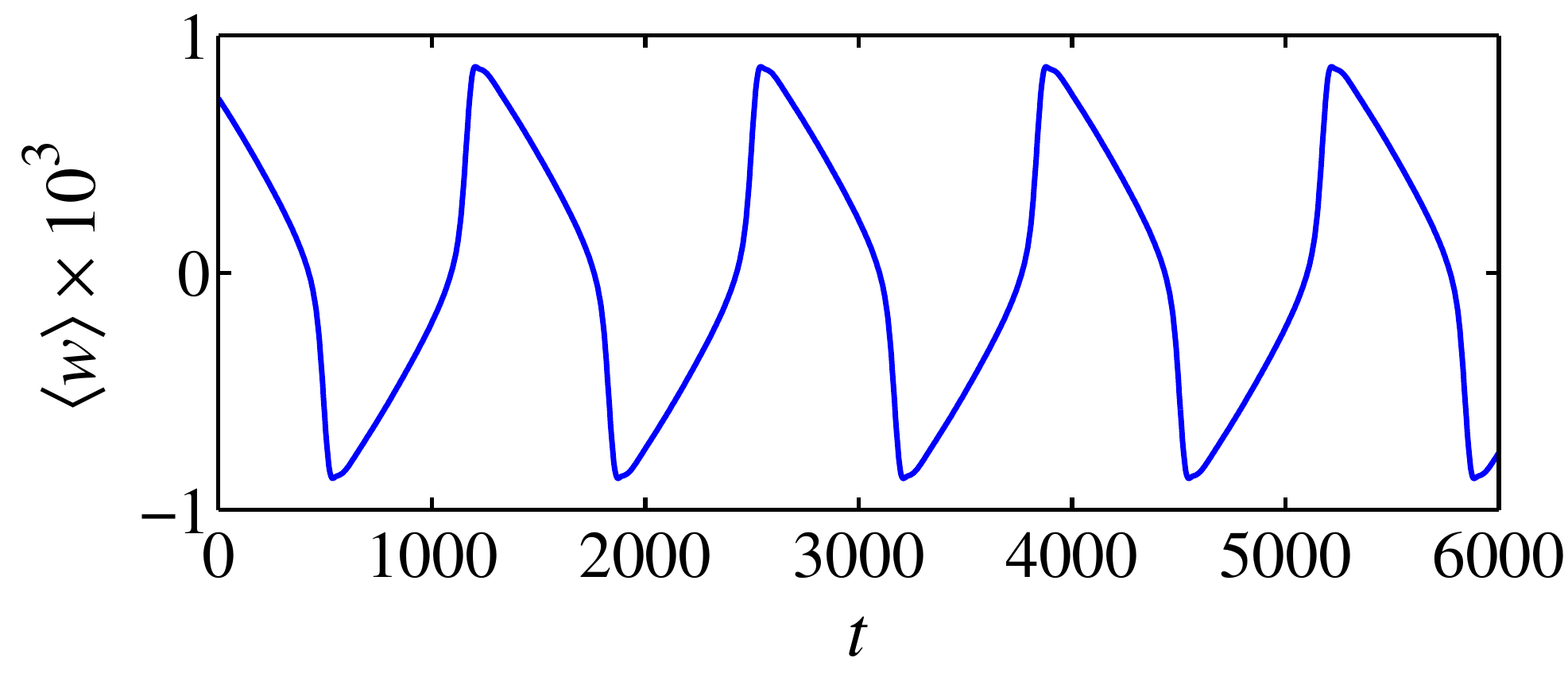}
\begin{picture}(0,0)
\put(-195,158){(\textit{a})}
\put(-195,75){(\textit{b})}
\end{picture}
\caption{\label{fig:6pi_lr_Ecf} \small Time series for the left-right shifting state (LR) at $L_x=6\pi$: (\textit{a}) cross-flow energy
$E_{\mathrm{cf}}$; (\textit{b}) mean spanwise velocity $\langle w \rangle$.}
\end{figure}

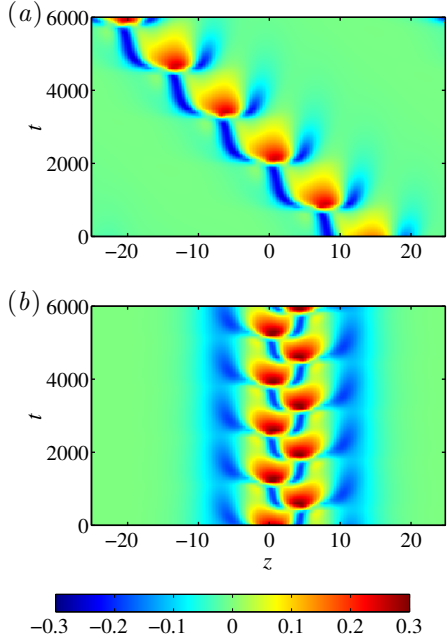
\begin{figure}
\centering
\includegraphics[scale=0.33]{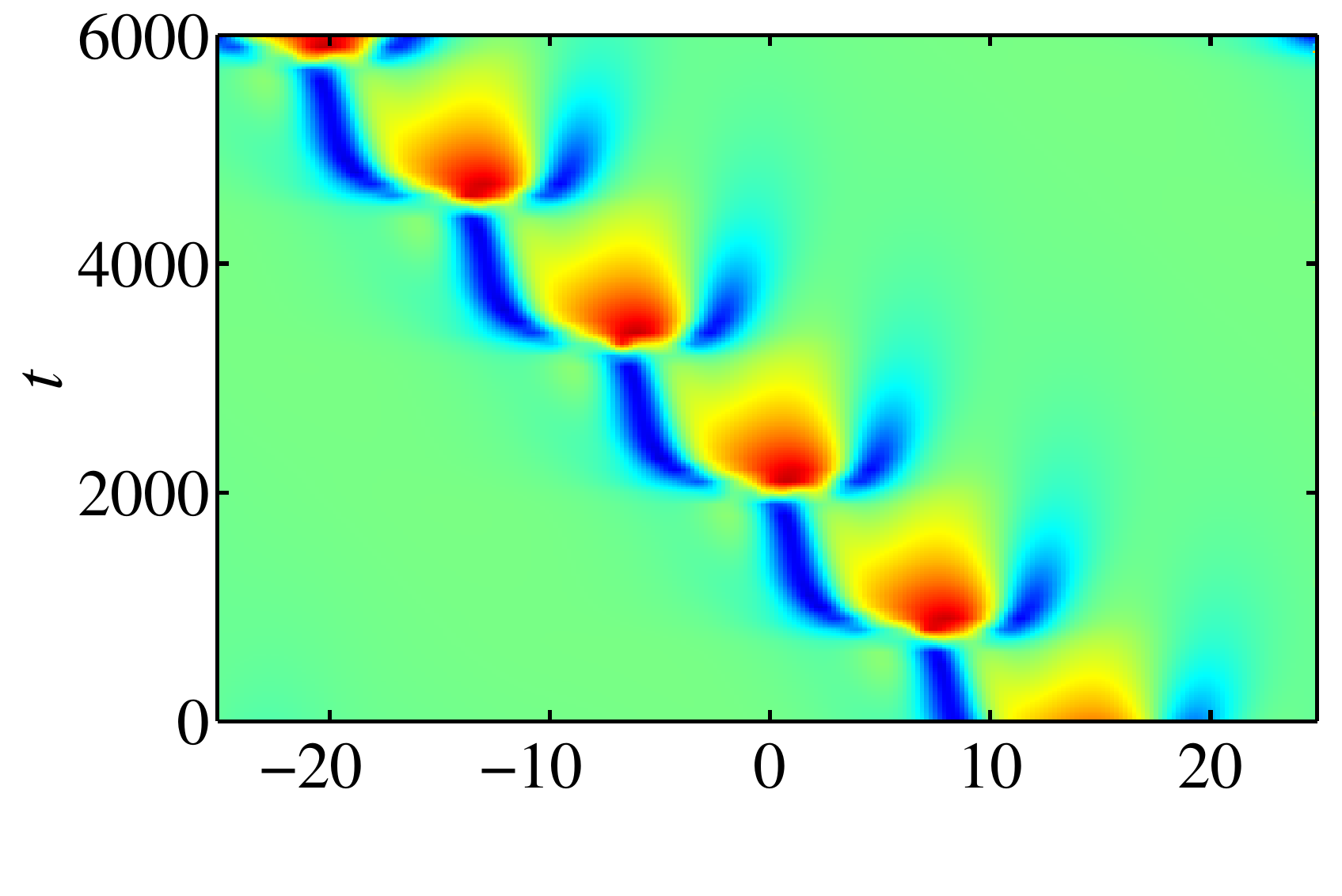}
\includegraphics[scale=0.33]{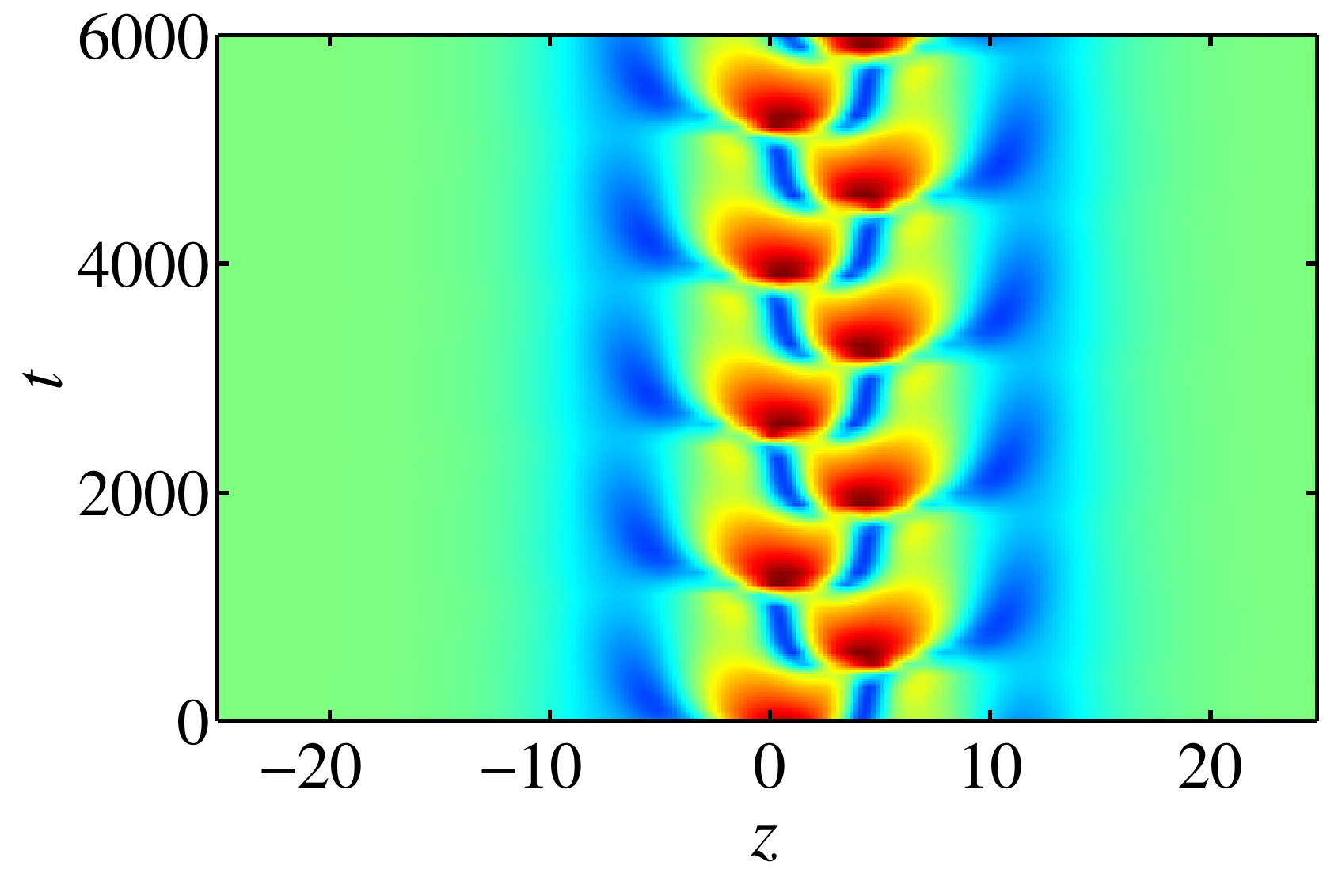}
\includegraphics[scale=0.33]{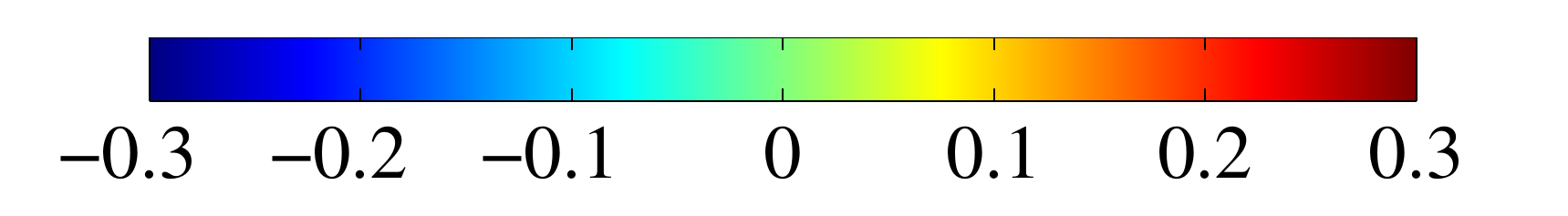}
\begin{picture}(0,0)
\put(-170,235){(\textit{a})}
\put(-170,126){(\textit{b})}
\end{picture}
\caption{\label{fig:6pi_st} \small Space--time diagrams of the streamwise velocity fluctuations $u'$ averaged in $x$ at $y=1$ for (\textit{a}) the
state
shifting left (L) and (\textit{b}) the state shifting repeatedly left and right (LR). For the online version: from blue to red, negative
(low-speed streaks) to positive (high-speed streaks) values of $u'$.}
\end{figure}

From here on, the Reynolds number is held fixed at $Re=500$. The edge state in the numerical domain of size \\*  $(L_x, L_y, L_z)=(6\pi, 15, 50)$ has
been discussed in detail in ref.~\cite{khapko_kreilos_schlatter_duguet_eckhardt_henningson_2013}, and we  briefly summarise the main findings here. In
this set-up three distinct edge states were found (modulo translational symmetries in $x$ and $z$), two of which are related by the symmetry $z
\rightarrow -z$. They all are localised in the spanwise direction and their dynamics is exactly periodic in time. The time evolution of the
cross-flow energy $E_{\mathrm{cf}}$ for
the states is shown in figs.~\ref{fig:6pi_l_Ecf}(\textit{a}) and~\ref{fig:6pi_lr_Ecf}(\textit{a}). This quantity can be considered a measure for
the amplitude of streamwise vortices and is defined as:
\begin{equation}
E_{\mathrm{cf}} = \frac 1{L_x L_z} \int_{\Omega} (v'^2 + w'^2)\, \mathrm dx\, \mathrm dy\, \mathrm dz \ ,
\end{equation}
where $v'$ and $w'$ are the wall-normal and spanwise velocity perturbations to the laminar solution $(U,V,W)$ and $\Omega$ stands for the
computational domain. For all three states the time signal alternates between calm regions with relatively low energy and bursts where significantly
higher values in energy are realised. The temporal dynamics as well as the spatial
structure of the edge states is shown in fig.~\ref{fig:6pi_st}
using a space--time diagram for the streamwise velocity perturbations $u'$ evaluated at $y=1$ and averaged in the $x$ direction. During the calm phase
the states consist of a pair of active high- and low-speed streaks with a slowly decaying low-speed streak on the side. After the burst of the
cross-flow energy the structure is destroyed but rapidly reforms modulo a shift in the spanwise direction. Based on the direction of the translation
we distinguish between the state alternatively shifting left and right (LR), and the ones shifting constantly in one direction, left (L, $z<0$) or
right (R, $z>0$). The information about the translation direction can be extracted from the time series of the spatially-averaged spanwise velocity
$\langle w \rangle= \int_0^{L_y} \hat{w}_{(0,0)}\, \mathrm dy$ displayed in figs.~\ref{fig:6pi_l_Ecf}(\textit{b}) and~\ref{fig:6pi_lr_Ecf}(\textit{b}), where $\hat{w}_{(0,0)}$ is the $(0,0)$ mode of the Fourier discretisation in the
horizontal plane at a given $y$ position. Here $\langle w
\rangle$ serves as a linear observable. Positive and negative values of $\langle w \rangle$ near the cross-flow energy peaks indicate shifts to the
right and left, respectively.

\begin{figure}
\centering
\includegraphics[width=.9\linewidth]{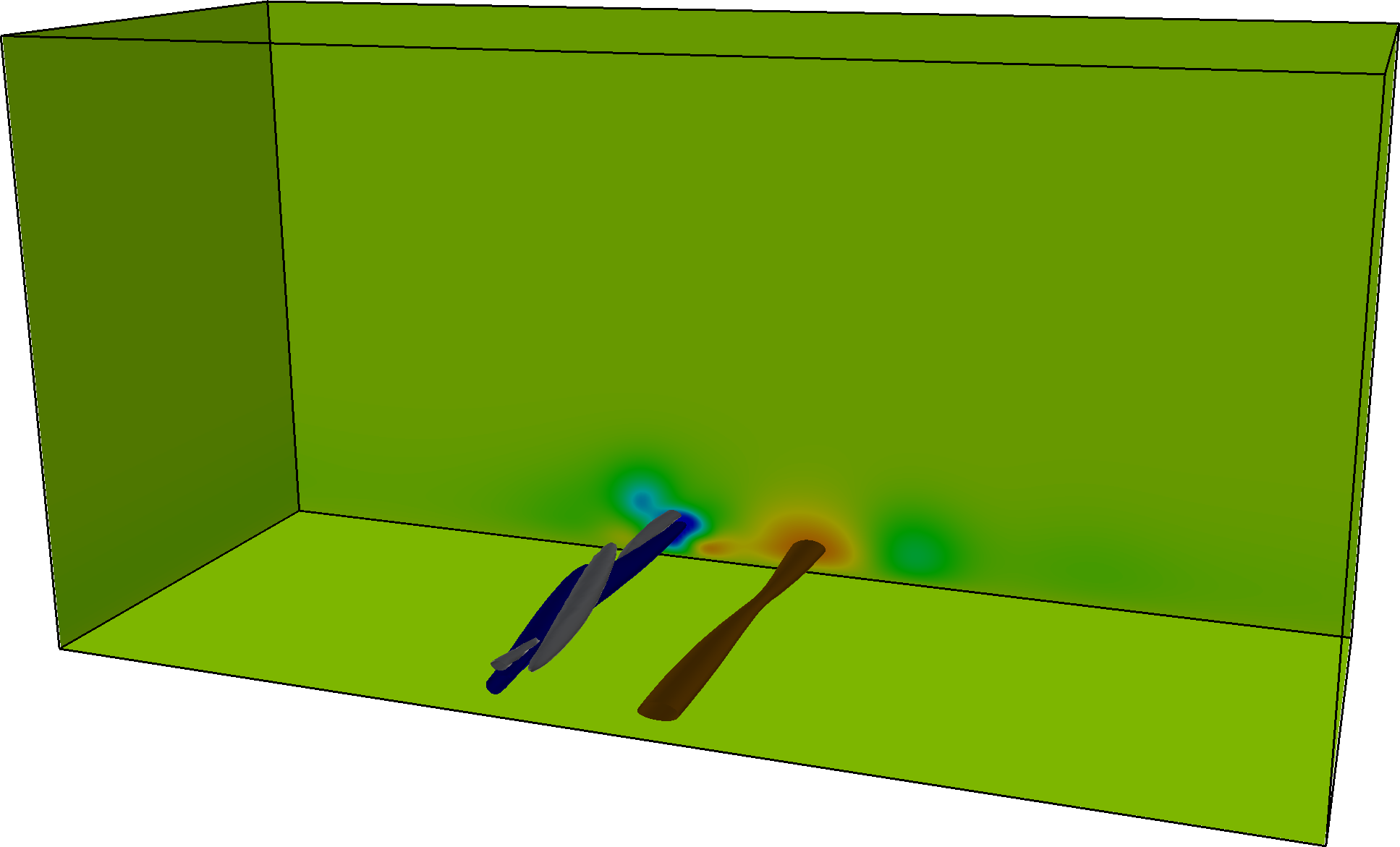}
\caption{\label{fig:6pi_structure} \small Snapshot of the L state for $L_x=6\pi$ shortly before a burst. Isocontours are $u' = -0.2$ in blue, $u'=0.1$
in red and $\lambda_2=-0.001$ in grey. A pair of counter-rotating vortices can be identified leaning over the low-speed streak. On the right, a
high-speed streak can be seen, the remnant of the preceding  burst. Flow from lower left to upper right.}
\end{figure}

The dynamics during one period of the cycle is similar for all three states, being equivalent to the one in the small domain
\cite{kreilos_veble_schneider_eckhardt_2013}, and bears strong resemblance with the self-sustaining cycle of wall turbulence described in
ref.~\cite{hamilton_kim_waleffe_1995}. In the beginning of the calm phase the state consists of almost streamwise-independent streamwise
streaks, with one
of the low-speed streaks accompanied by a pair of quasi-streamwise vortices. The
vortices
increase in strength and size, further bending the streak (see fig.~\ref{fig:6pi_structure}). Eventually the
low-speed streak breaks up, which
corresponds to
the burst in the cross-flow energy. During the breakdown process streamwise vortices are re-created
in the vicinity of the destroyed low-speed streak.
They lead to the creation of a high-speed streak at the same spanwise position with two low-speed streaks on the sides, and the loop is closed (see
also ref.~\cite{khapko_kreilos_schlatter_duguet_eckhardt_henningson_2013}). Thus,
compared to the cycle from \cite{hamilton_kim_waleffe_1995} there is an additional spatial
aspect, manifesting itself in shifts of the structure in the spanwise direction.

\begin{figure}
\centering
\includegraphics[scale=0.33]{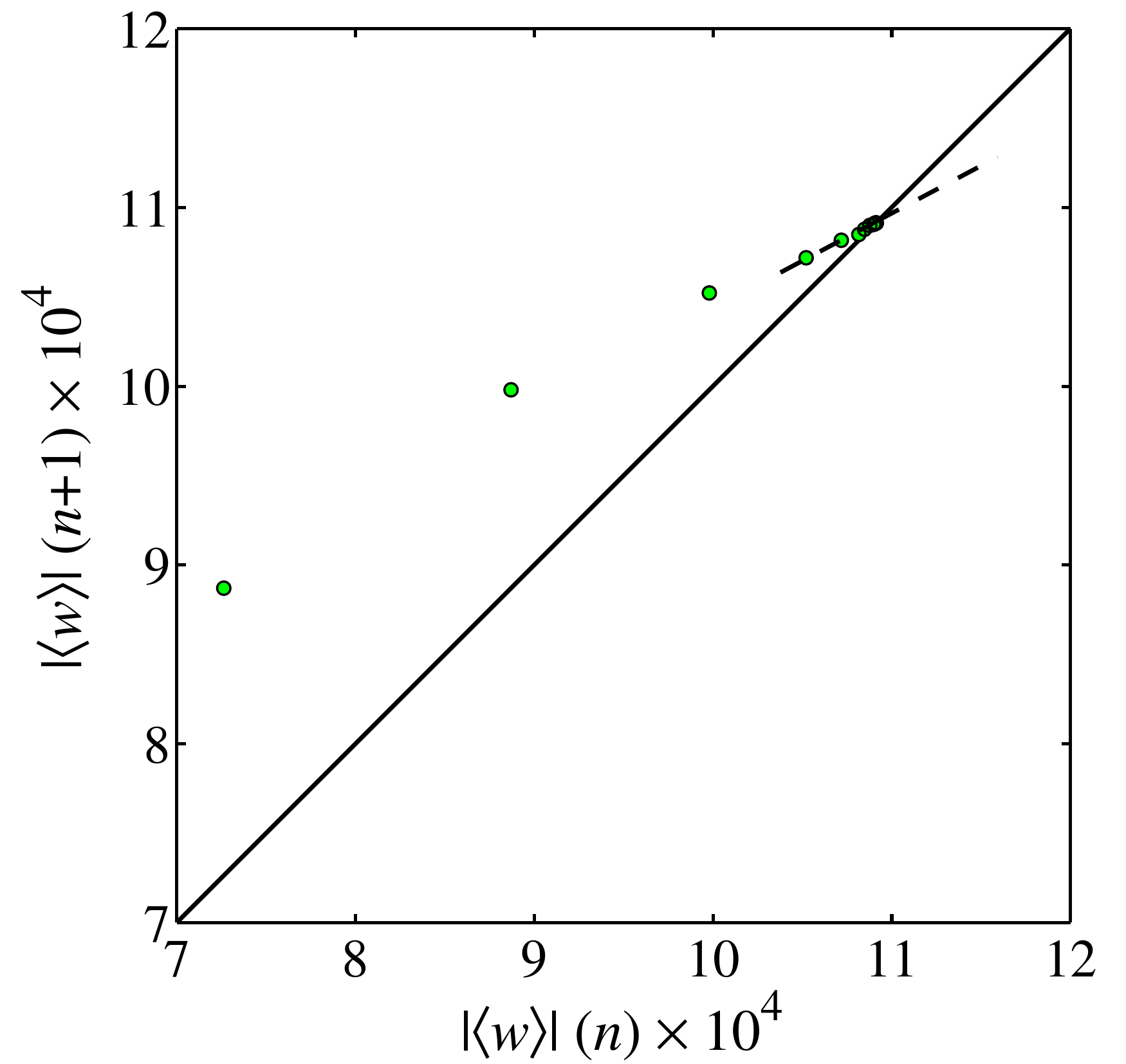}
\caption{\label{fig:6pi_ret} \small First return map  of the absolute value of mean spanwise velocity $\abs{\langle w \rangle}$ at the peaks of the
cross-flow energy $E_{\mathrm{cf}}$ for the L state at $L_x=6\pi$. The slope $\beta$ of the approach to the diagonal, indicated by the dashed and
solid lines, respectively, is close to $0.5$, meaning that the state is stable.}
\end{figure}

The linear stability of the edge states with respect to perturbations within the basin boundary can be checked by considering return maps of
$\abs{\langle w \rangle}$ sampled at times corresponding to the maxima of $E_{\mathrm{cf}}$. In such a representation a period-$k$ orbit corresponds
to $k$ different points on the diagonal of the $k$-th return map. The slope $\beta$ near these points in the return map should be equal and indicates
the linear stability properties of the orbit: $|\beta|>1$ indicates instability whereas $|\beta|<1$ indicates stability. From fig.~\ref{fig:6pi_ret},
which includes the approach to the L state in the first return map, we can for instance deduce that $\beta \approx 0.5$ and that the periodic orbit is
hence a stable attractor on the edge.

\subsection{Bifurcation diagram}

When the length $L_x$ of the computational domain is reduced to $4\pi$, only few edge trajectories converge to a periodic LR state while the
others stay erratic even for very long edge tracking times. In particular, we never got convergence to a periodic state when initiating the
bisection with random initial conditions. Despite being erratic the state remains strongly localised in $z$ for all times, with the active part
consisting of a pair of low- and high-speed streaks. As in the periodic case, high-$E_{\mathrm{cf}}$ bursts are followed by spanwise
translations of
the whole structure. However, the time between the bursts is not constant and the shifts vary unpredictably in direction and distance. Understanding
and characterising the bifurcations connecting the periodic to the erratic regime is the focus of this study.

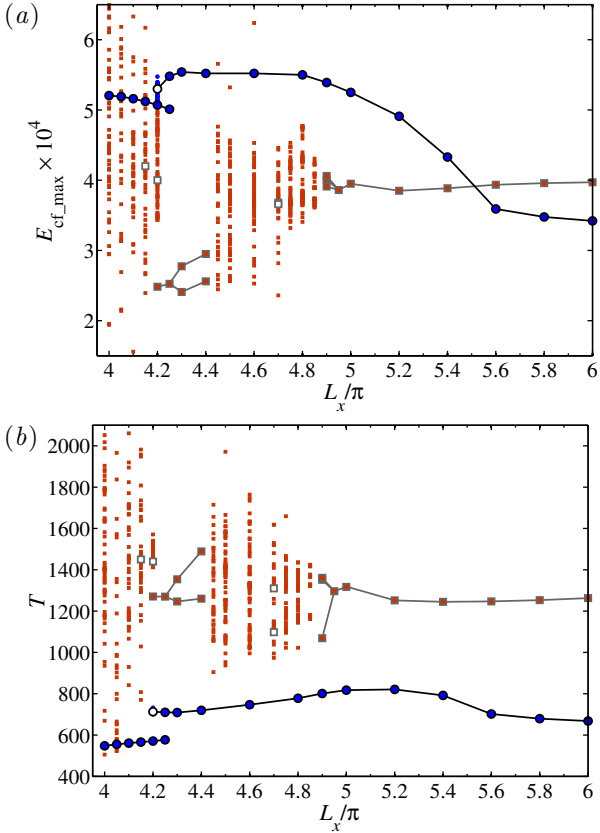
\begin{figure}[h]
\centering
\includegraphics[scale=0.33]{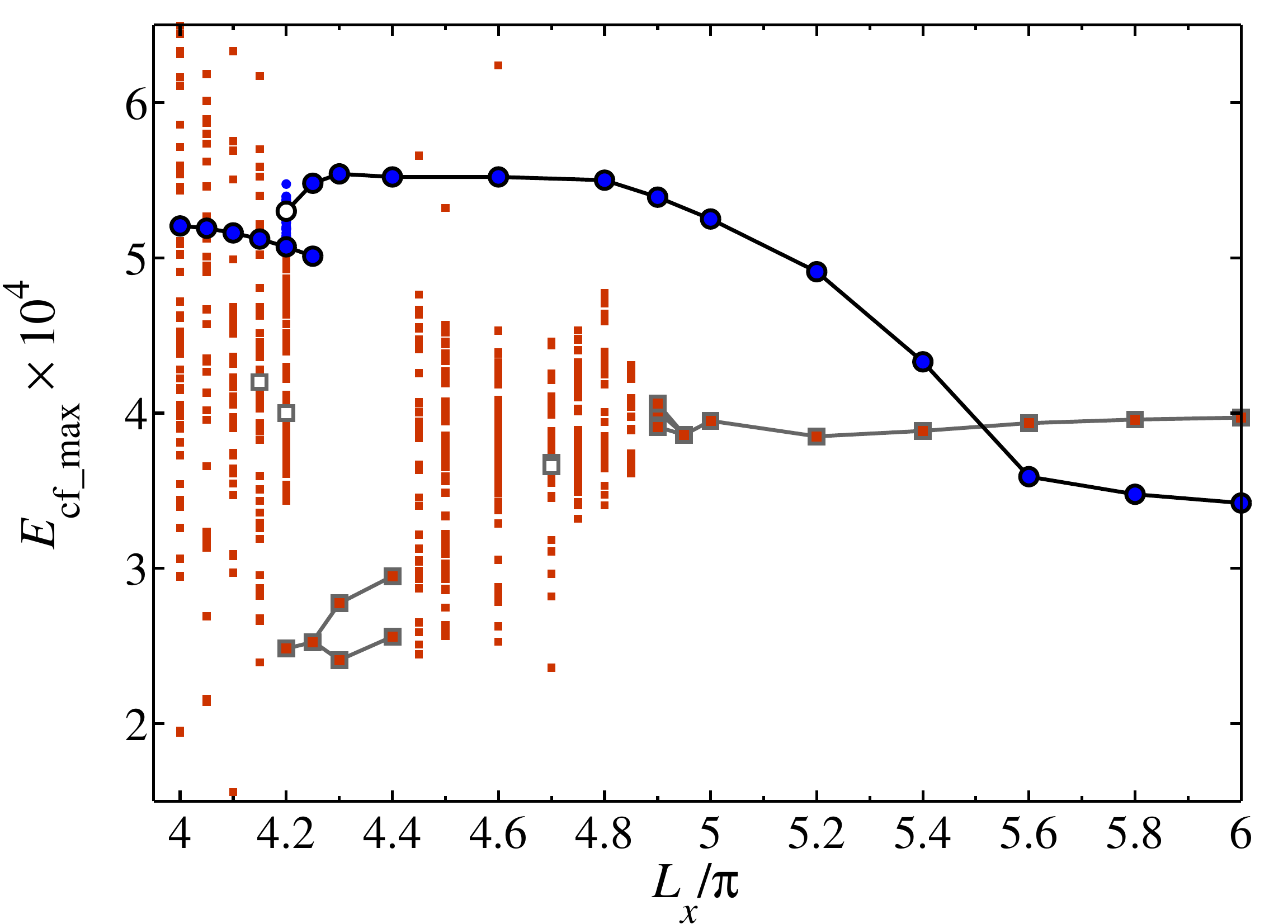}
\includegraphics[scale=0.33]{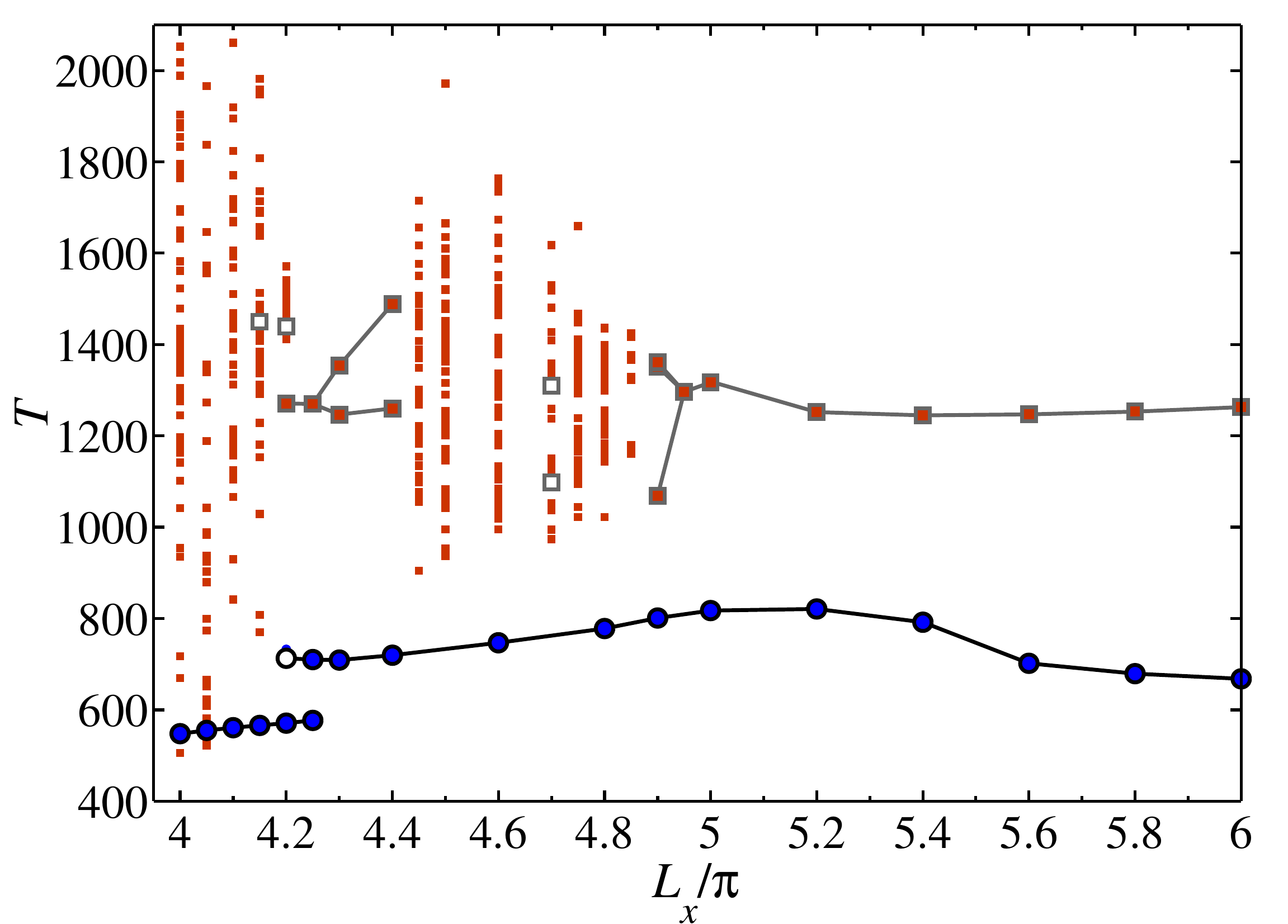}
\begin{picture}(0,0)
\put(-230,305){(\textit{a})}
\put(-230,146){(\textit{b})}
\end{picture}
\caption{\label{fig:bif_Lx} \small Bifurcation diagram in terms of (\textit{a}) cross-flow energy peaks $E_{\mathrm{cf\_max}}$ and (\textit{b})
inter-burst
periods $T$. The branches for the L and LR states are represented with squares (red online) and circles (blue online), respectively. Larger symbols
represent stable (filled) and unstable (empty) states, whereas erratic behaviour is denoted with smaller
symbols.}
\end{figure}

The long temporal period of the obtained states together with the large number of degrees of freedom involved, makes numerical continuation not
feasible with the tools at hand. Therefore, we chose to vary the parameter $L_x$ in discrete steps, starting from $L_x=6\pi$. Since L and R states are
the same under $z\rightarrow -z$ transformation, it is sufficient to focus on one of them only. Starting from $L_x=6\pi$ the instantaneous flow fields
of both L and LR states were shrunk to shorter domains. The resulting fields were used as initial conditions for the edge tracking, which was
performed until the edge state in the considered domain was found. Initially, $L_x$ was decreased from $6\pi$ to $4\pi$ in steps of $0.2\pi$. As soon
as qualitative changes were observed at some length, the same
procedure was
performed from the last periodic state with smaller step size. Decreases in $L_x$ were regularly complemented with increases
in $L_x$ in order to check for
hysteresis. Finally the important bifurcation regions were re-sampled using a finer resolution of $0.05 \pi$ in $L_x$. The simulations altogether
represent a total of nearly $10^6$ CPU hours.

\begin{figure}[t]
\centering
\includegraphics[scale=0.33]{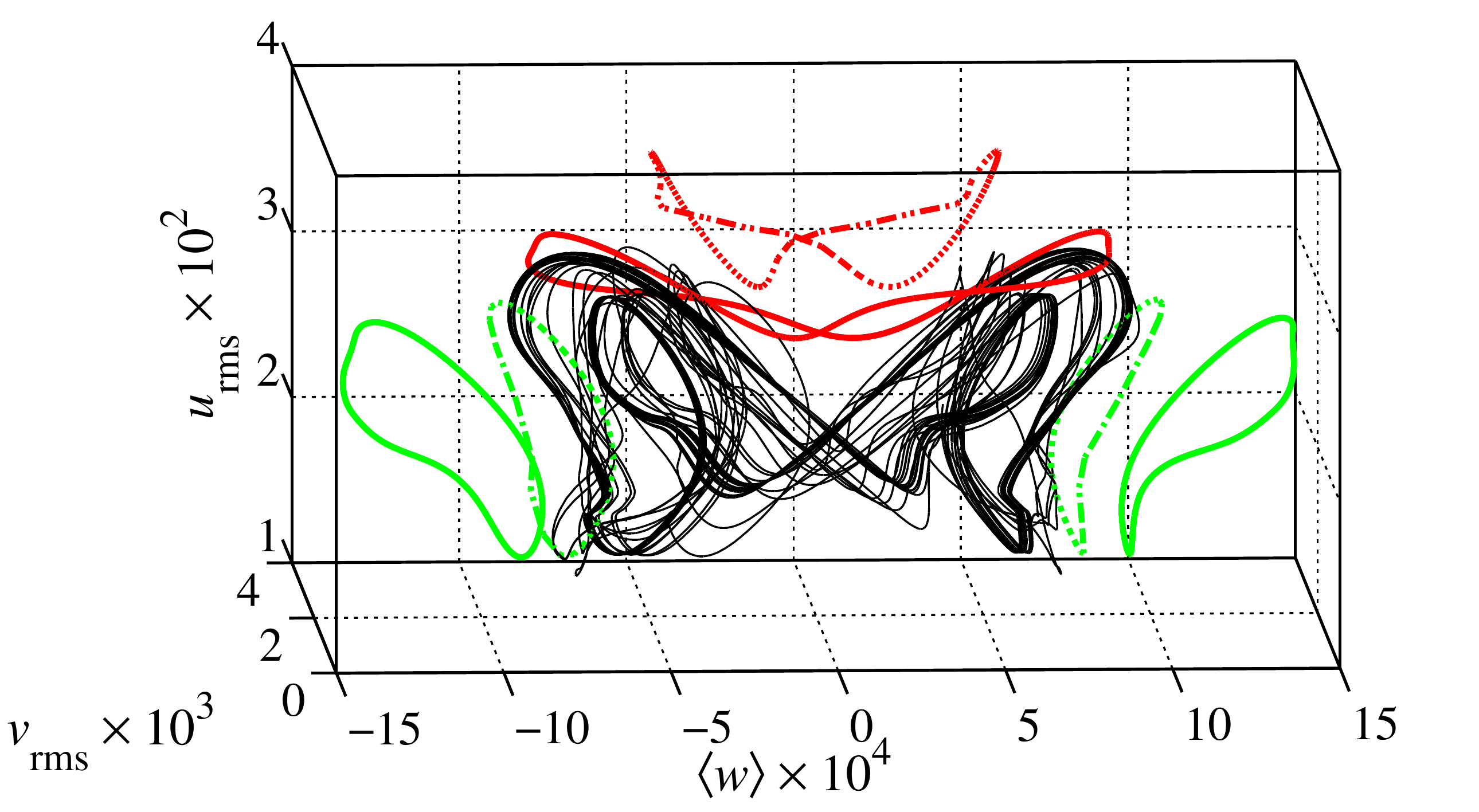}
\includegraphics[scale=0.33]{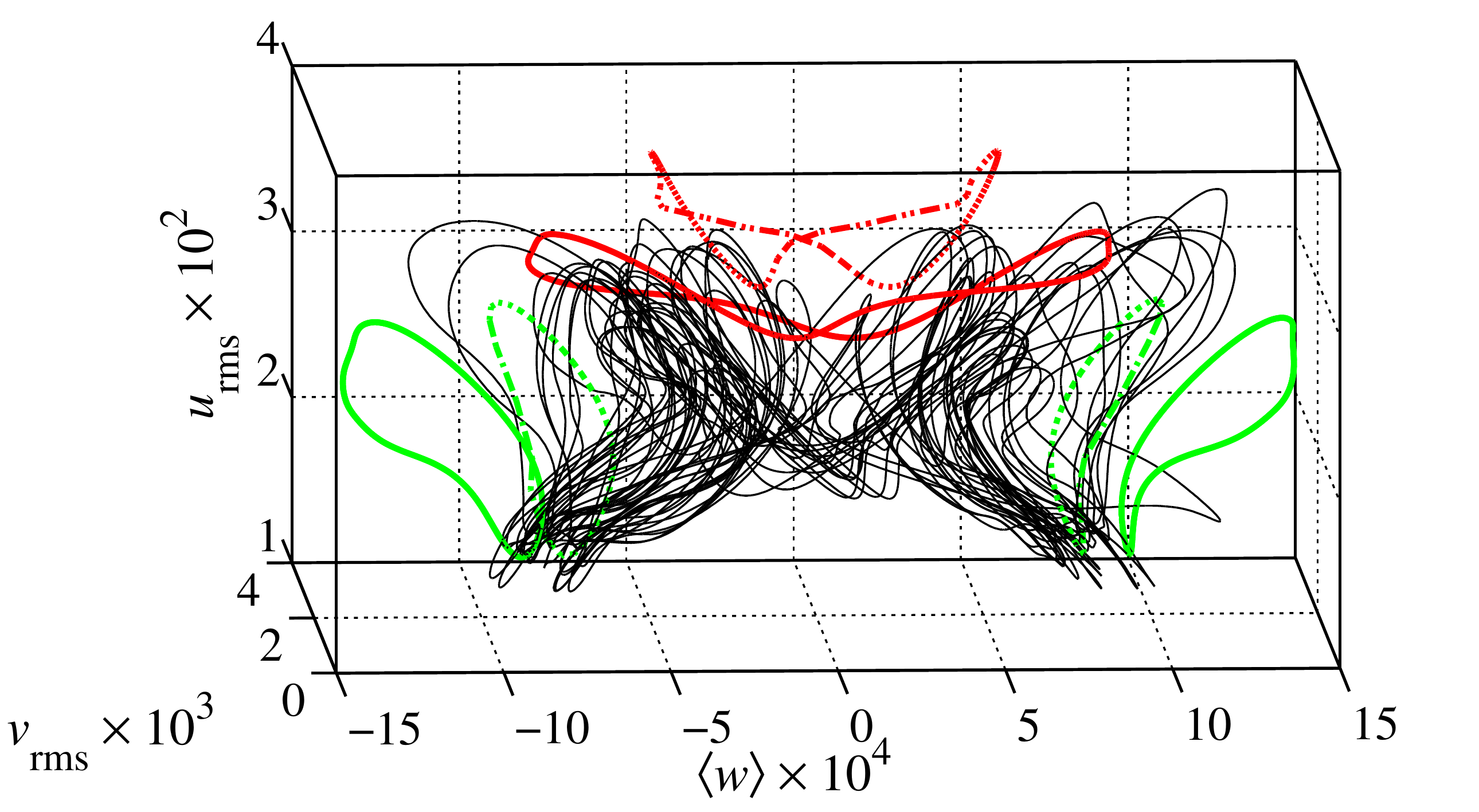}
\caption{\label{fig:phase-space} \small Representation of the dynamics in a space spanned by the mean spanwise velocity $\langle w \rangle$ and the
root-mean-square velocity fluctuations in the wall-normal ($v_{\mathrm{rms}}$) and streamwise ($u_{\mathrm{rms}}$) components for length $L_x=4.7\pi$
and $L_x=4.5\pi$ in (\textit{a}) and (\textit{b}), respectively. The dynamics on the relative attractor within the edge (shown as a full line
{\color{black}\solid}) is intermittent in case (\textit{a}) and erratic in case (\textit{b}). In addition, we show the trajectories of the periodic LR
state at $L_x=6\pi$ ({\color{red}\solid})
and $L_x=4.6 \pi$ ({\color{red}\dotdashed}), as well as that of the $L$ and $R$ states at $L_x=6\pi$ ({\color{green}\solid}) and $L_x=5\pi$
({\color{green}\dotdashed}).}
\end{figure}

The results obtained using this procedure are displayed in fig.~\ref{fig:bif_Lx} via two bifurcation diagrams. In fig.~\ref{fig:bif_Lx}(\textit{a}),
the maxima of $E_{\mathrm{cf}}$ are plotted for each value of $L_x$. We obtain two independent branches associated with L (squares) and LR (circles)
states. Stable periodic states are represented by large filled symbols. If the state is $n$-periodic it corresponds to $n$ different points in the
diagram. Erratic states are denoted with smaller symbols. Unequal amount of points in different cases is due to varying length of the trajectories on
the edge and irregular periods between two consecutive $E_{\mathrm{cf}}$ peaks. Nonetheless, each of the erratic simulations is an order of magnitude
longer than the average time it takes to reach a periodic state, with up to $100,000$ time units in the longest chaotic simulations. If some range of
values are frequently visited by the chaotic trajectory the corresponding state in the diagram is marked with a large empty
symbol. In
fig.~\ref{fig:bif_Lx}(\textit{b}) an alternative representation is shown, where the time intervals $T$ between consecutive peaks of $E_{\mathrm{cf}}$
are considered instead. Some of the states identified within the bifurcation diagram are displayed in phase-space projections
($u_{\mathrm{rms}}$,$v_{\mathrm{rms}}$,$\langle w \rangle$), see fig.~\ref{fig:phase-space}. The intervals in $L_x$ chosen there contain the
parameters where the aperiodic orbits are calculated and indicate how the orbits move with $L_x$. The aperiodic trajectories meander through phase
space, transiently approach the loci of the L and R states associated to larger values of $L_x$ and move between them along segments of the LR
trajectory.

While the modulation in cross-flow energy of the LR state changes with $L_x$, the orbit stays exactly periodic until $L_x=4.25\pi$.
At $L_x=4.2\pi$ it becomes weakly unstable, and the trajectory spends considerable time in its vicinity before leaving and
exploring
different part of the phase space. For some initial conditions, the localised state begins drifting in one direction (as described in the next
paragraph). Some initial conditions lead to a new branch of LR states, characterised by a shorter period $T$. This new LR-branch has been tracked down
to as low as $L_x=3.6\pi$, below which periodicity in time is lost. It can also be continued to higher values of $L_x \geq 4.25\pi$, however we did
not track the branch further. Whether it reconnects with the original LR branch remains an open question. This new state was not identified in
ref.~\cite{khapko_kreilos_schlatter_duguet_eckhardt_henningson_2013}, which suggests that it has a smaller basin of attraction than the original LR
state.

The stability range for the L state is narrower compared to the LR state, as it does not extend below $L_x=4.95\pi$. For the next value we probed,
$L_x=4.9\pi$, a longer but stable $3$-period state emerges as relative attractor on the edge, the full
period of which consists of three consecutive shifts in
each direction. We denote it by $\mathrm{L}^3\mathrm{R}^3$ (note that in fig.~\ref{fig:bif_Lx} two of the three symbols are very
close to each other). Between $4.85\pi$ and $4.75\pi$ the dynamics is erratic in terms of the time evolution of the cross-flow energy, however
repeating the same pattern of translations in the spanwise direction, with four consecutive shifts in each direction for $L_x=4.85\pi$,
three consecutive
shifts in each direction for $L_x=4.8\pi$ and irregular combinations between two and three shifts in each direction for
$L_x=4.75\pi$. The
corresponding trajectories show no indication of convergence. Thus, even if the underlying orbits are stable, considerably longer
simulation times may be required for convergence in this case. For $L_x=4.7\pi$ Pomeau--Manneville intermittency is observed in the form of chaotic dynamics alternating with visits to an
unstable $\mathrm{L}^2\mathrm{R}^2$ orbit. Below this value the dynamics is erratic, except in a stable period-$1$ and a period-$2$ window between
$4.2\pi$ and $4.4\pi$, where an L state is recovered again. Additionally, an intermittently chaotic L state is obtained from the LR-branch at
$4.2\pi$ and on the L-branch at $4.15\pi$. It is marked with empty squares for the corresponding values of $L_x$ in the bifurcation diagram.
Due to the interesting dynamics on this branch, some parameter values will be discussed in more detail below.

\subsection{Investigation of the L-branch}

\subsubsection{Period-$3$ state ($L_x=4.9\pi$)}

\begin{figure}
\centering
\includegraphics[scale=0.33]{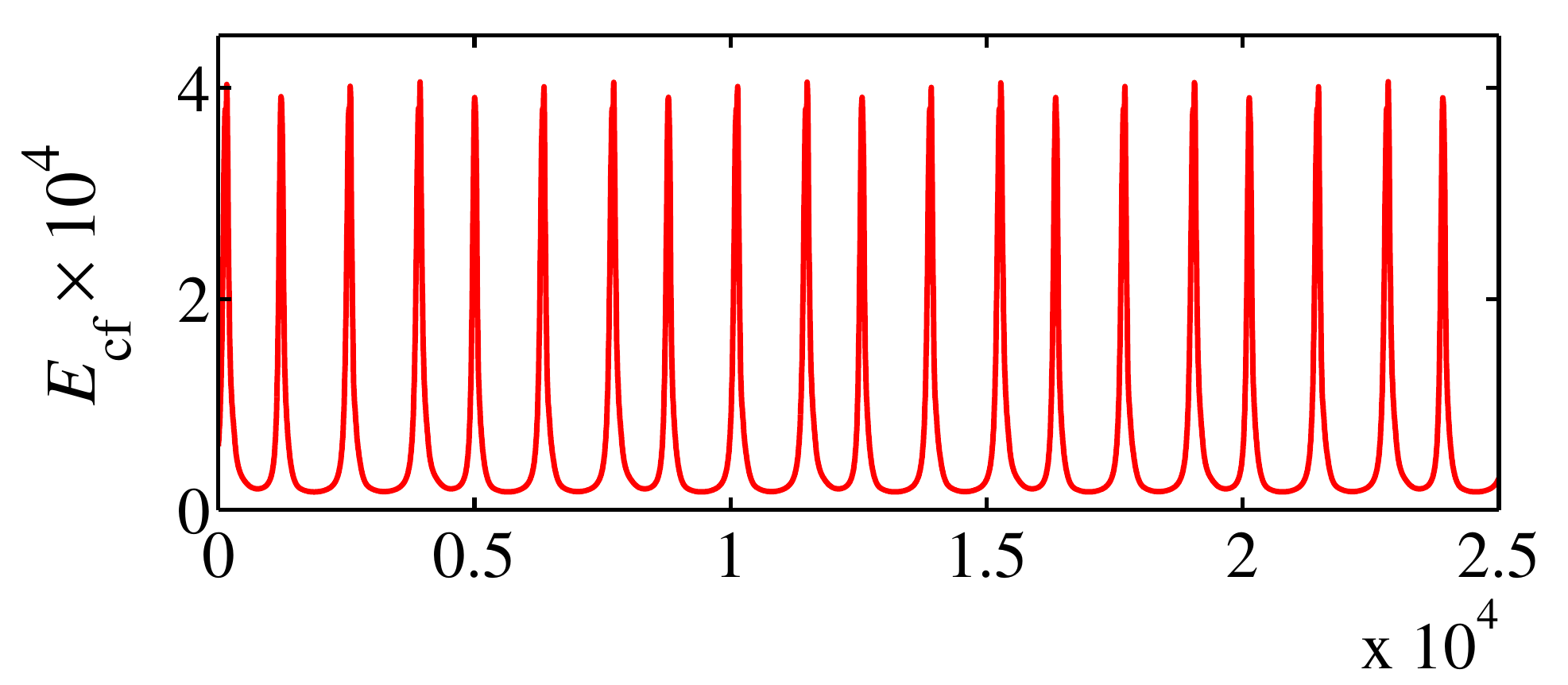}
\includegraphics[scale=0.33]{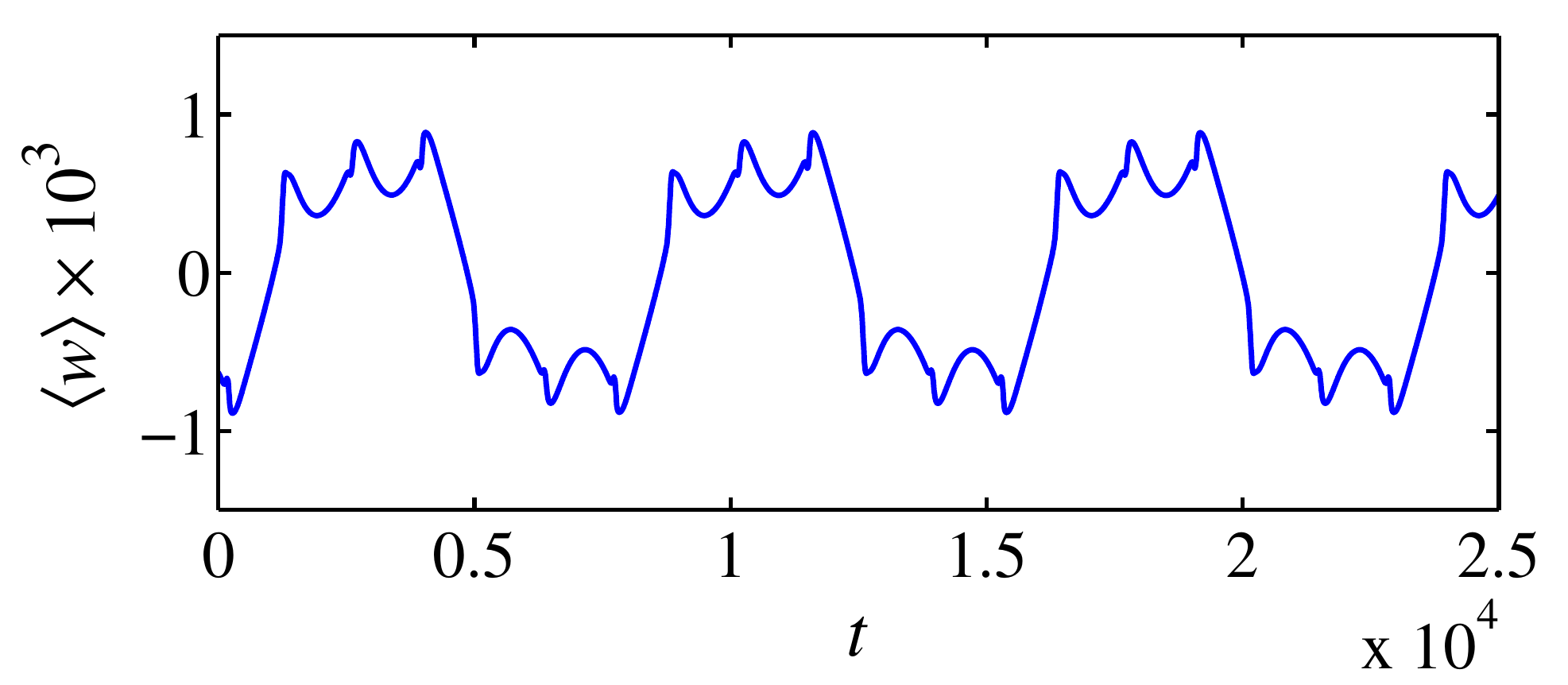}
\begin{picture}(0,0)
\put(-185,158){(\textit{a})}
\put(-185,75){(\textit{b})}
\end{picture}
\caption{\label{fig:4.9pi_Ecf} \small Time series for the $3$-periodic $\mathrm{L}^3\mathrm{R}^3$ state at \mbox{$L_x=4.9\pi$}: (\textit{a})
cross-flow
energy $E_{\mathrm{cf}}$; (\textit{b}) mean spanwise velocity $\langle w \rangle$.}
\end{figure}

\begin{figure*}
\centering
\includegraphics[scale=0.33]{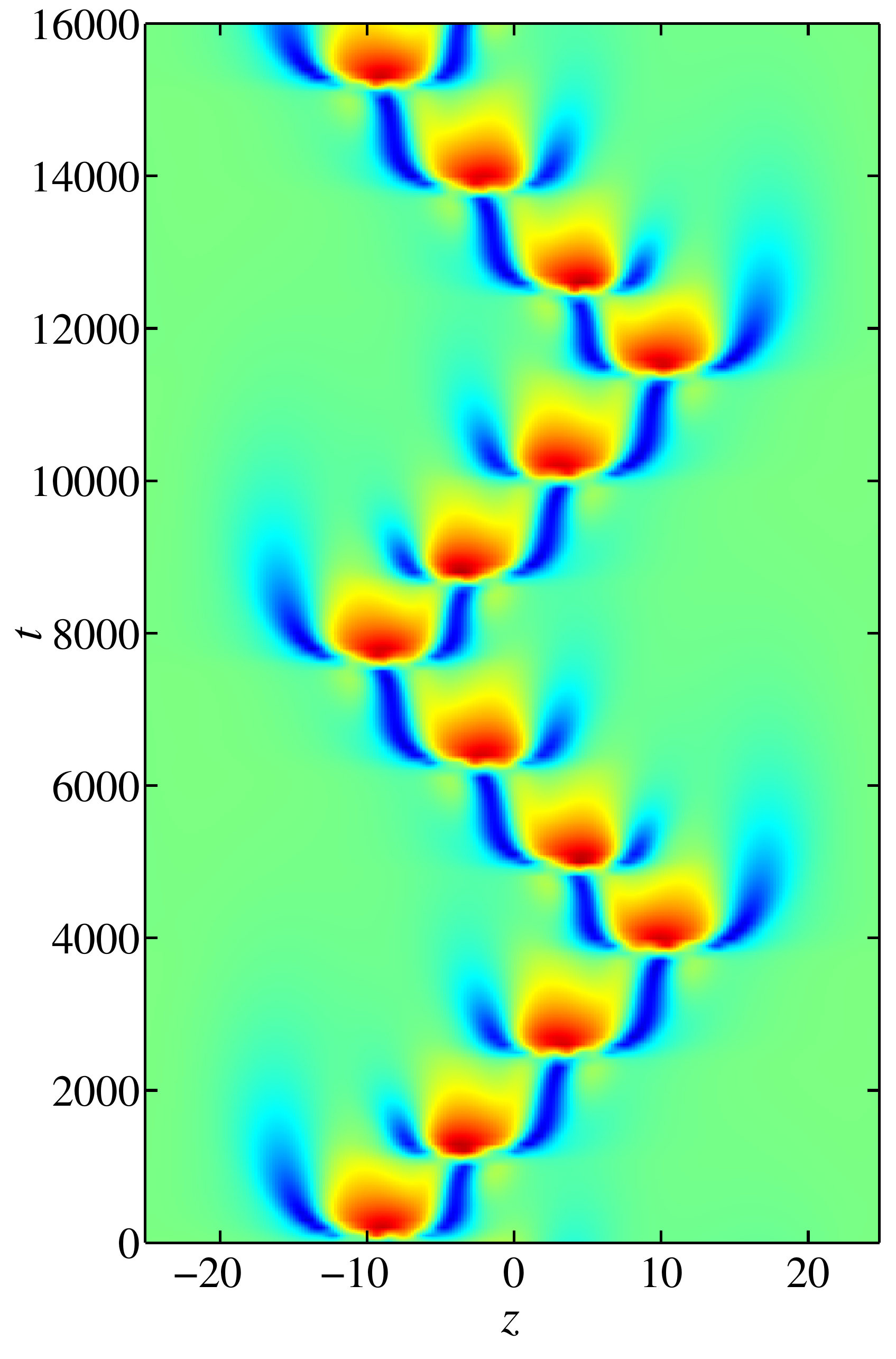}
\includegraphics[scale=0.33]{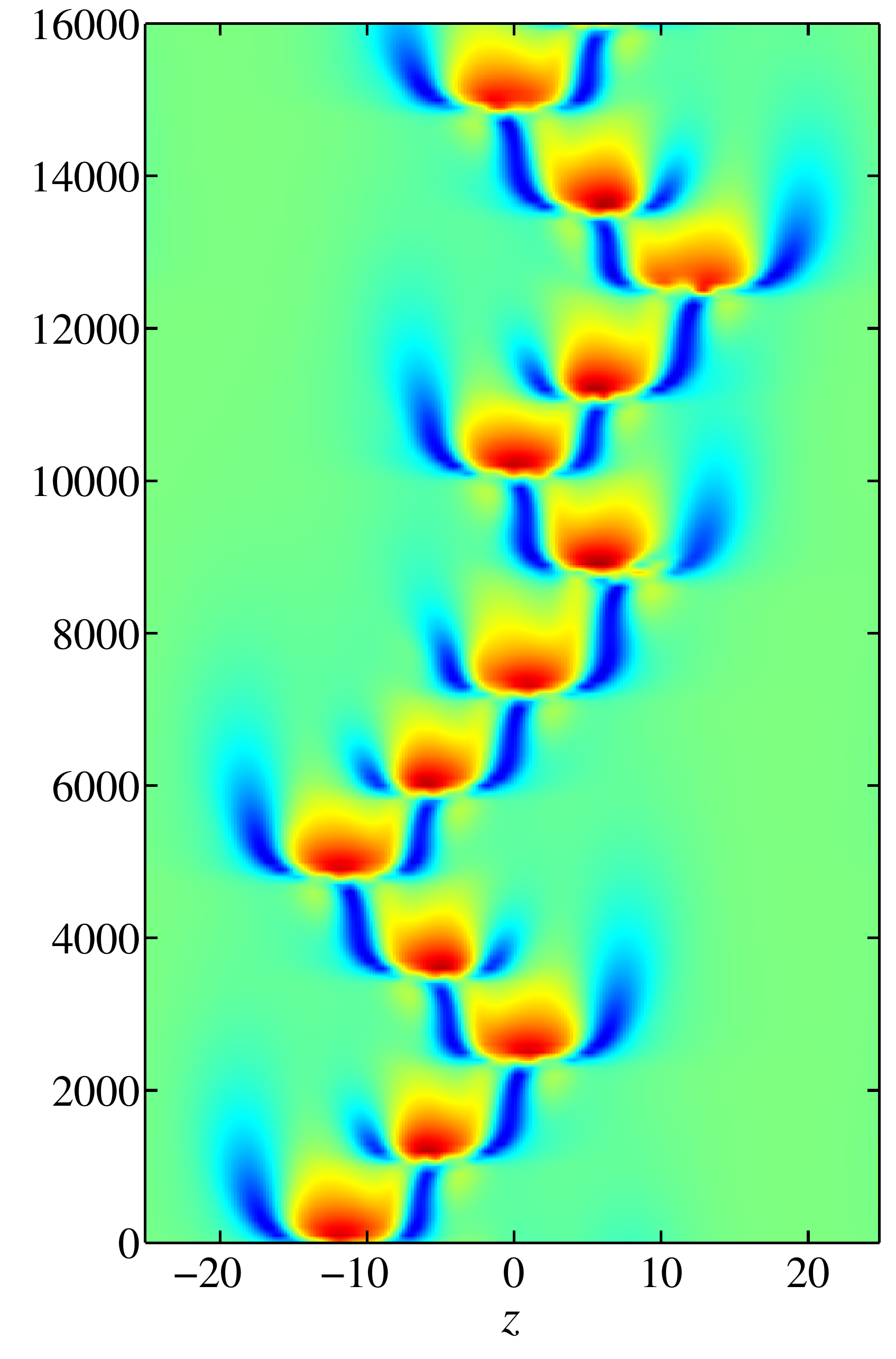}
\includegraphics[scale=0.33]{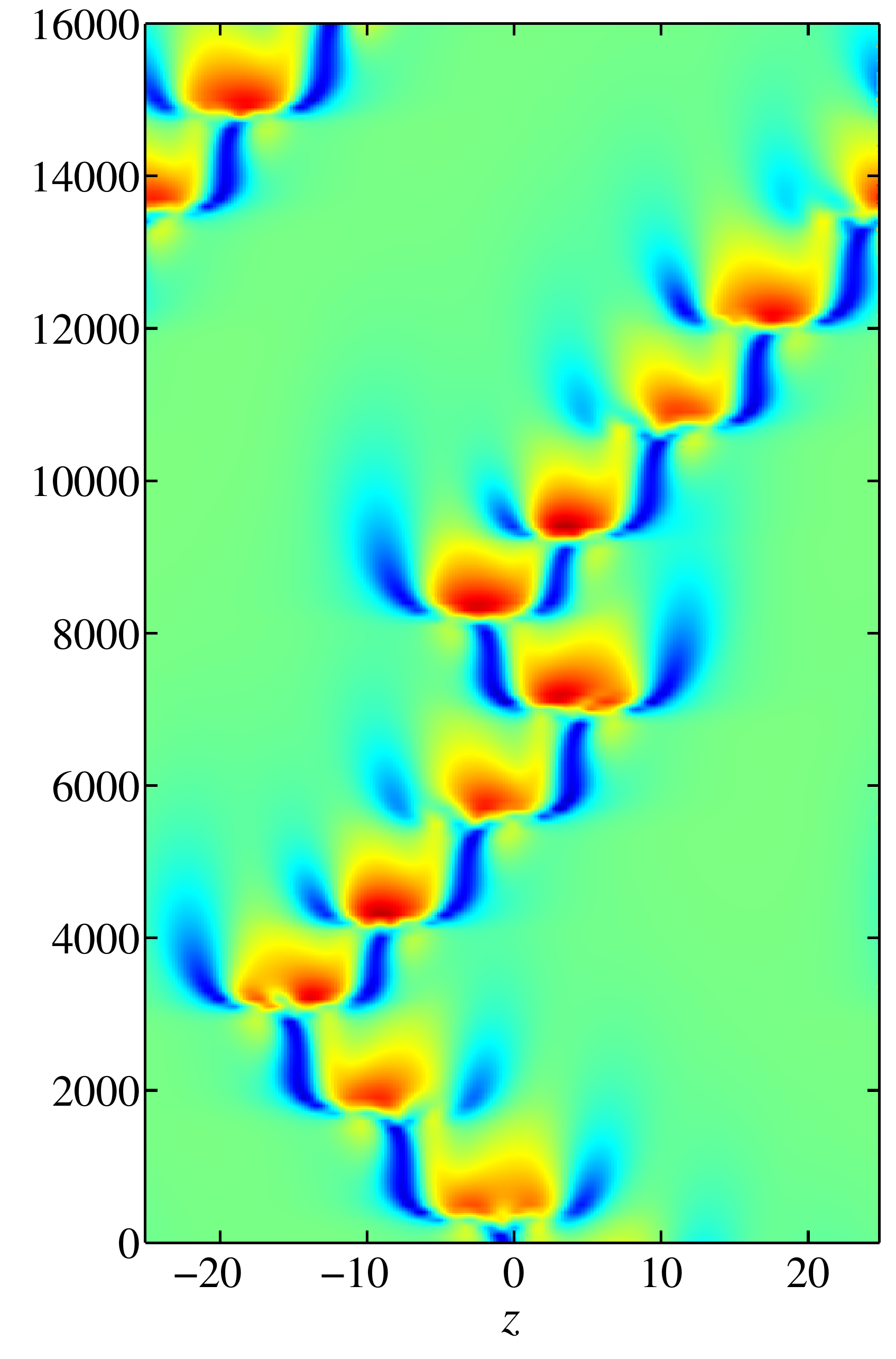}
\begin{picture}(0,0)
\put(-495,225){(\textit{a})}
\put(-329,225){(\textit{b})}
\put(-164,225){(\textit{c})}
\end{picture}
\caption{\small Space--time diagrams of the streamwise velocity fluctuations $u'$ averaged in $x$ at $y=1$ for some states on the L-branch:
(\textit{a})
Period-$3$ state at $L_x=4.9\pi$ ($\mathrm{L}^3\mathrm{R}^3$); (\textit{b}) Intermittent state at $L_x=4.7\pi$; (\textit{c}) Chaotic state at
$L_x=4.5\pi$. The colourmap is the same as in fig.~\ref{fig:6pi_st}.}
\label{fig:long_st}
\end{figure*}
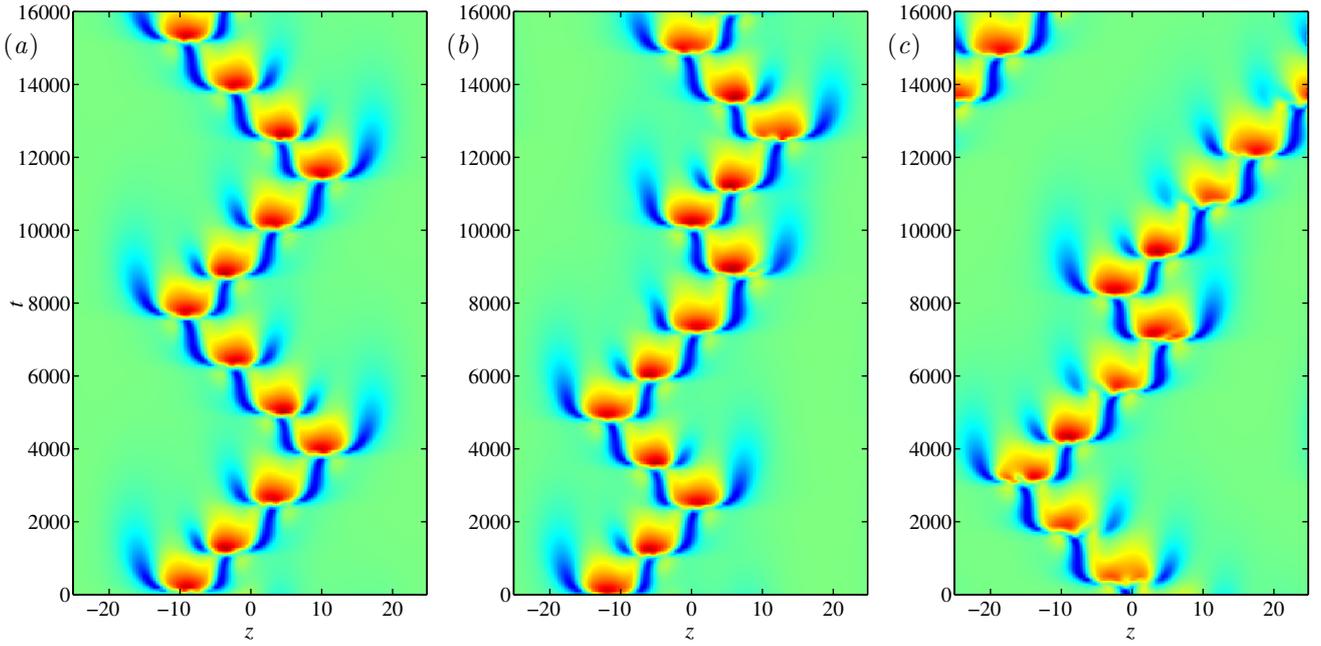

For $L_x=4.9\pi$ a longer periodic state is obtained. As is evident from fig.~\ref{fig:4.9pi_Ecf}, the state is $3$-periodic in terms of the time
evolution of $E_{\mathrm{cf}}$. Part of the space--time diagram for this state is shown in fig.~\ref{fig:long_st}(\textit{a}), where the state is
seen to alternate between three shifts in each direction. We thus denote this state as $\mathrm{L}^3\mathrm{R}^3$.

\subsubsection{Intermittent state ($L_x=4.7\pi$)}

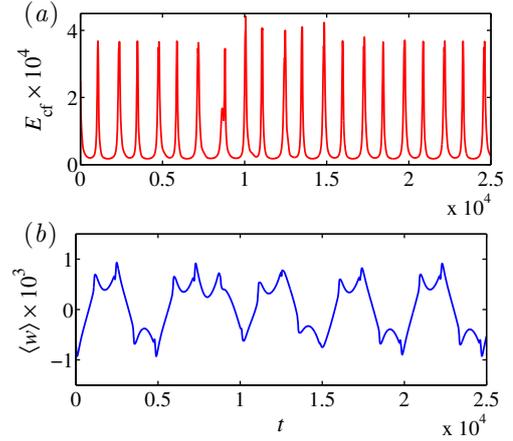
\begin{figure}
\centering
\includegraphics[scale=0.33]{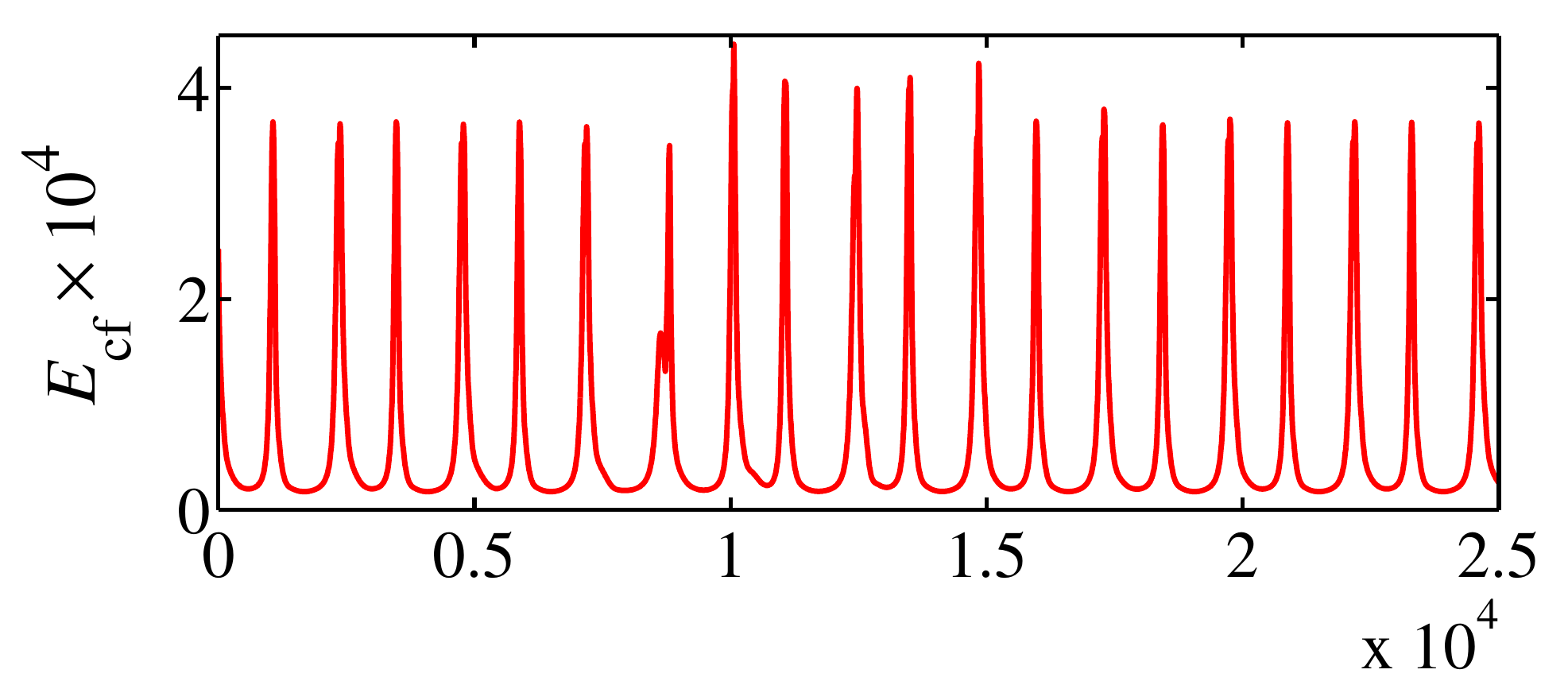}
\includegraphics[scale=0.33]{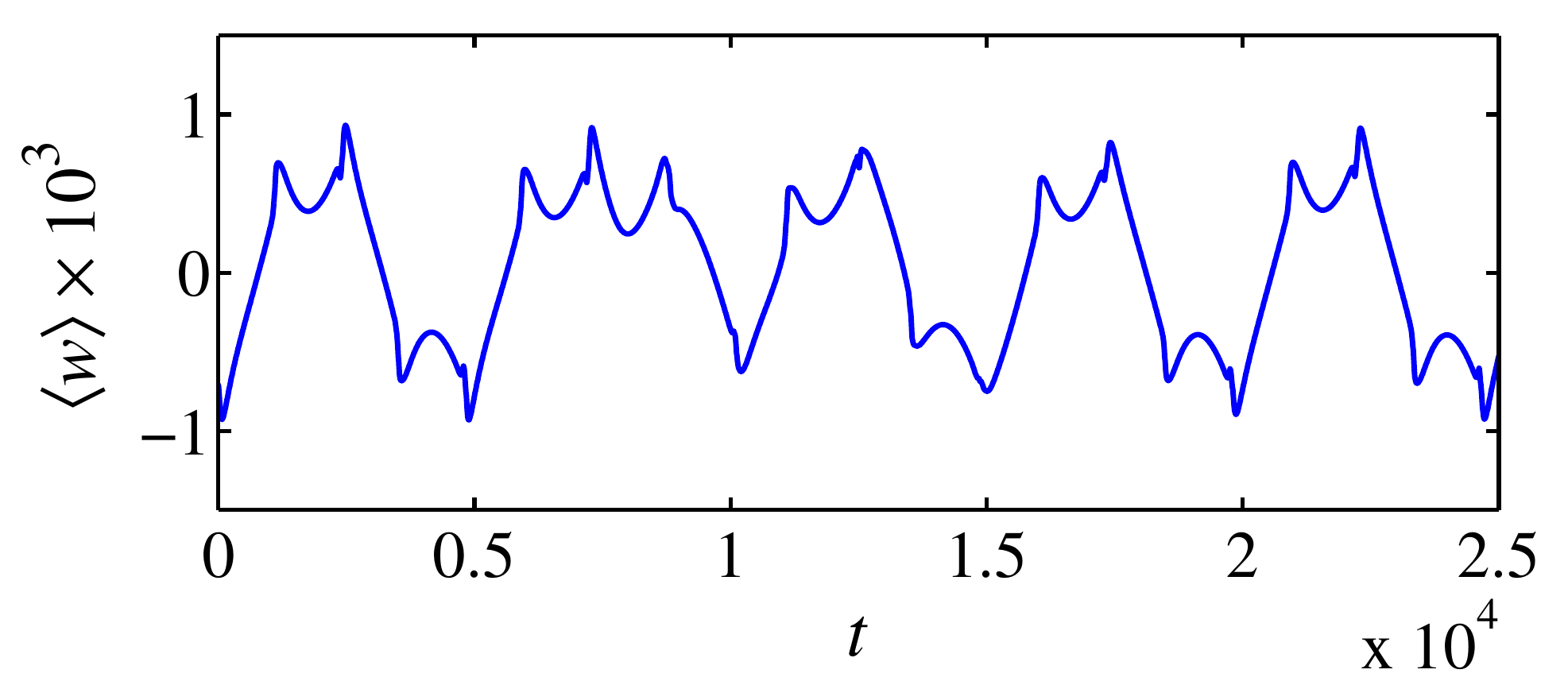}
\begin{picture}(0,0)
\put(-185,158){(\textit{a})}
\put(-185,75){(\textit{b})}
\end{picture}
\caption{\label{fig:4.7pi_Ecf} \small Time series for the intermittent state at $L_x=4.7\pi$: (\textit{a}) cross-flow energy $E_{\mathrm{cf}}$;
(\textit{b})
mean spanwise velocity $\langle w \rangle$.}
\end{figure}

\begin{figure}
\centering
\includegraphics[scale=0.33]{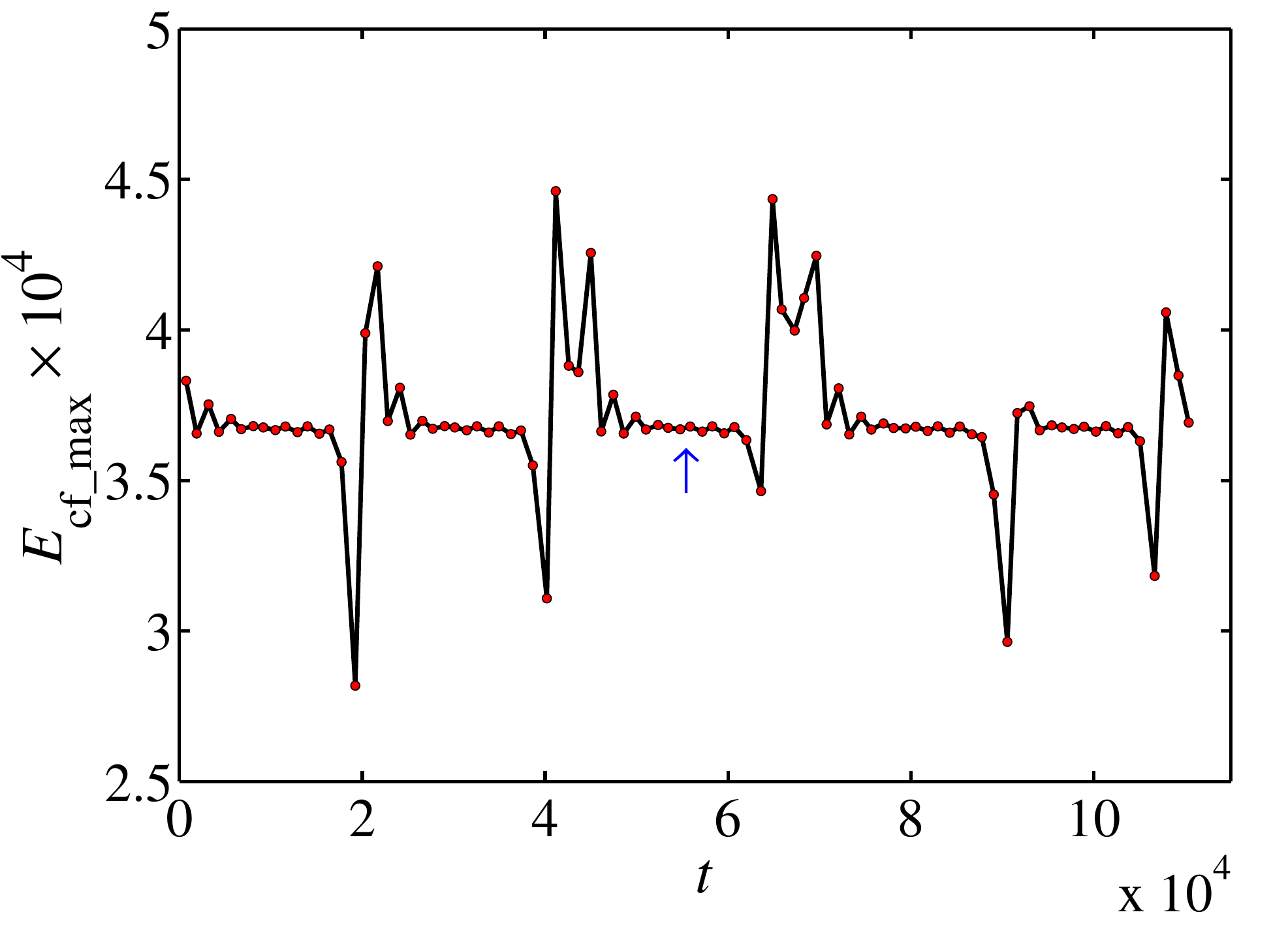}
\caption{\label{fig:4.7pi_Ecf_max} \small Cross-flow energy maxima $E_{\mathrm{cf\_max}}$ for the intermittent state at $L_x=4.7\pi$. The starting
time
for fig.~\ref{fig:long_st}(\textit{b}) and fig.~\ref{fig:4.7pi_Ecf} corresponds to $t=54,800$ in this representation, marked with the vertical
arrow in the figure.}
\end{figure}

\begin{figure}
\centering
\includegraphics[scale=0.33]{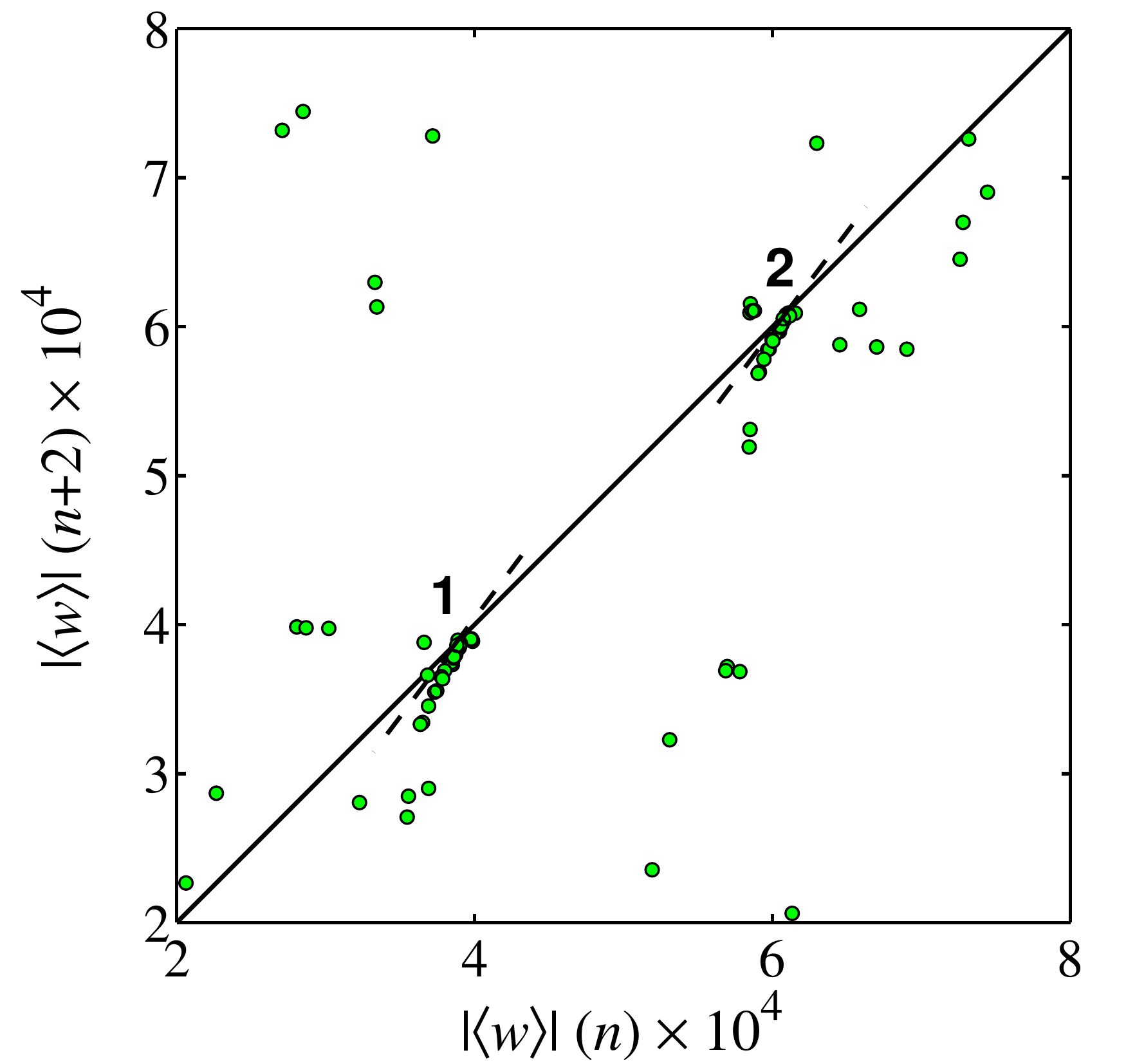}
\caption{\label{fig:4.7pi_ret} \small Second return map  of $\abs{\langle w \rangle}$ at the peaks of $E_{\mathrm{cf}}$ for the intermittent state at
$4.7\pi$. Both slopes, indicated by the dashed lines, are
approximately the same and are larger than $1$, with $\beta_1 \approx \beta_2 \approx 1.3$, proving that the $\mathrm{L}^2\mathrm{R}^2$ orbit is
unstable on the edge.}
\end{figure}

Lowering $L_x$ down to $4.7\pi$ on the L-branch, the periodic behaviour is lost. Part of the time evolution of $E_{\mathrm{cf}}$ and $\langle w
\rangle$ in this case is shown in fig.~\ref{fig:4.7pi_Ecf}. The dynamics seems to repeatedly spend considerable amount of time in the vicinity of the
state with two shifts in each direction ($\mathrm{L}^2\mathrm{R}^2$), before suddenly leaving and being quickly re-injected again. These transient
approaches to the $\mathrm{L}^2\mathrm{R}^2$ state can be clearly seen in fig.~\ref{fig:4.7pi_Ecf_max}, where the maxima of the cross-flow energy
$E_{\mathrm{cf\_max}}$ are plotted against time for the full trajectory. Again, in order to assess the stability of this $2$-period state the second
return map of $\abs{\langle w \rangle}$ is considered (see fig.~\ref{fig:4.7pi_ret}). Clearly, both slopes near the diagonal are larger than $1$,
meaning that the underlying $\mathrm{L}^2\mathrm{R}^2$ state exists at least for neighbouring parameters and is here not stable. A part of the
space--time diagram capturing this phenomena is shown in fig.~\ref{fig:long_st}(\textit{b}), where regular $\mathrm{L}^2\mathrm{R}^2$ shifts are
interrupted by short aperiodic motion. A phase-space representation of this state, together with some of the periodic states is shown in
fig.~\ref{fig:phase-space}(\textit{a}). Despite the chaotic motion, the underlying structure of the $\mathrm{L}^2\mathrm{R}^2$ orbit can be
identified.

This type of behaviour was first described by Pomeau and Manneville in $1980$ as a key element of transition to chaos through intermittency
\cite{pomeau_manneville_1980}. Depending on the bifurcation leading from periodic to intermittent dynamics different types of intermittency are
defined. In practice, in order to identify
the relevant intermittency scenario, converged statistics of the inter-burst times would be needed for a continuous range of values $L_x$ close to the bifurcation point,
which is too costly.

\subsubsection{Chaotic state ($L_x=4.5\pi$)}

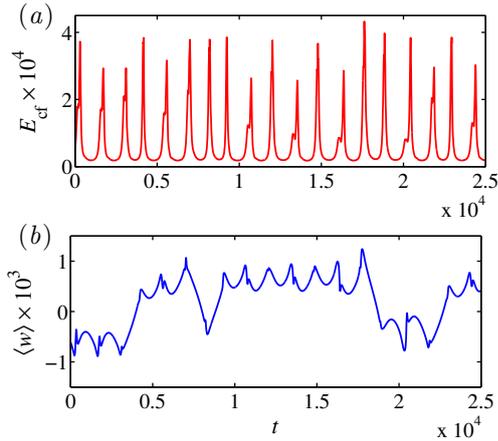
\begin{figure}
\centering
\includegraphics[scale=0.33]{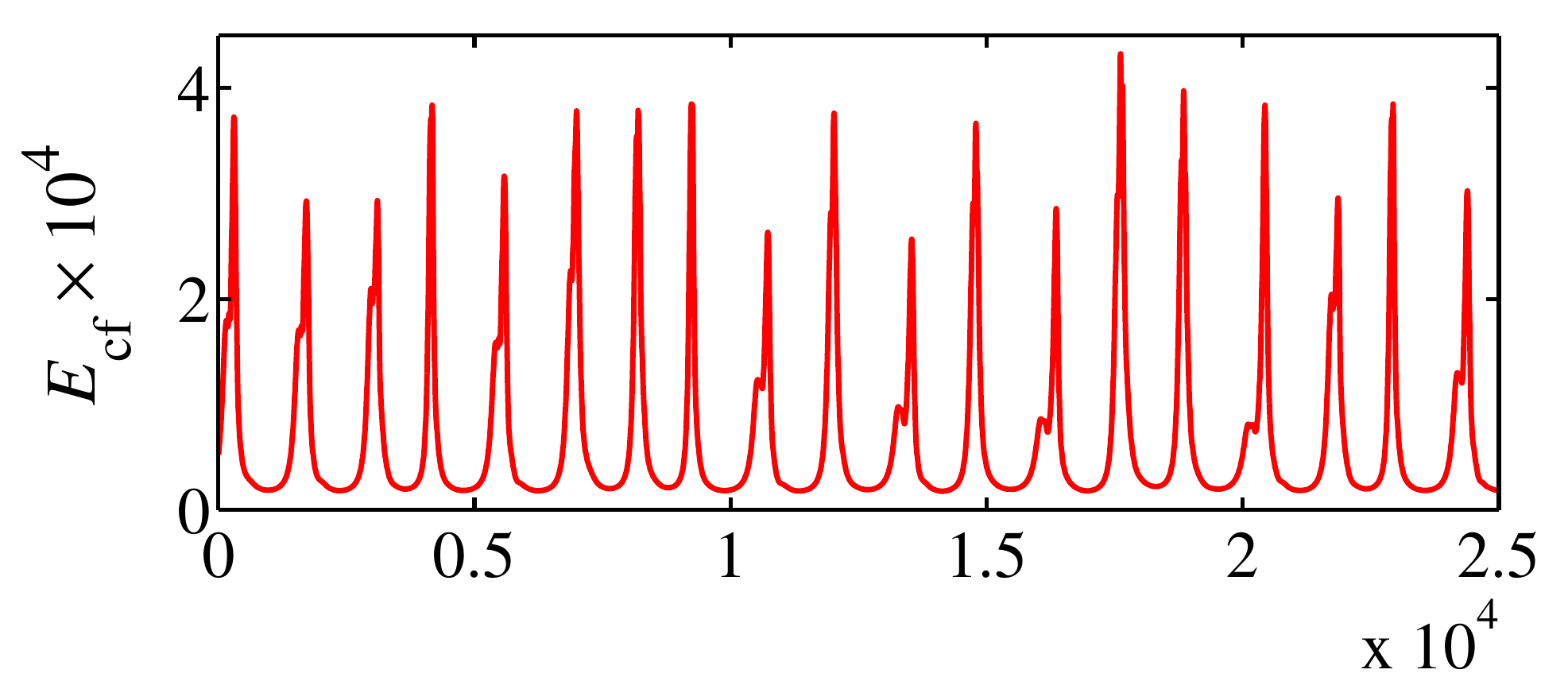}
\includegraphics[scale=0.33]{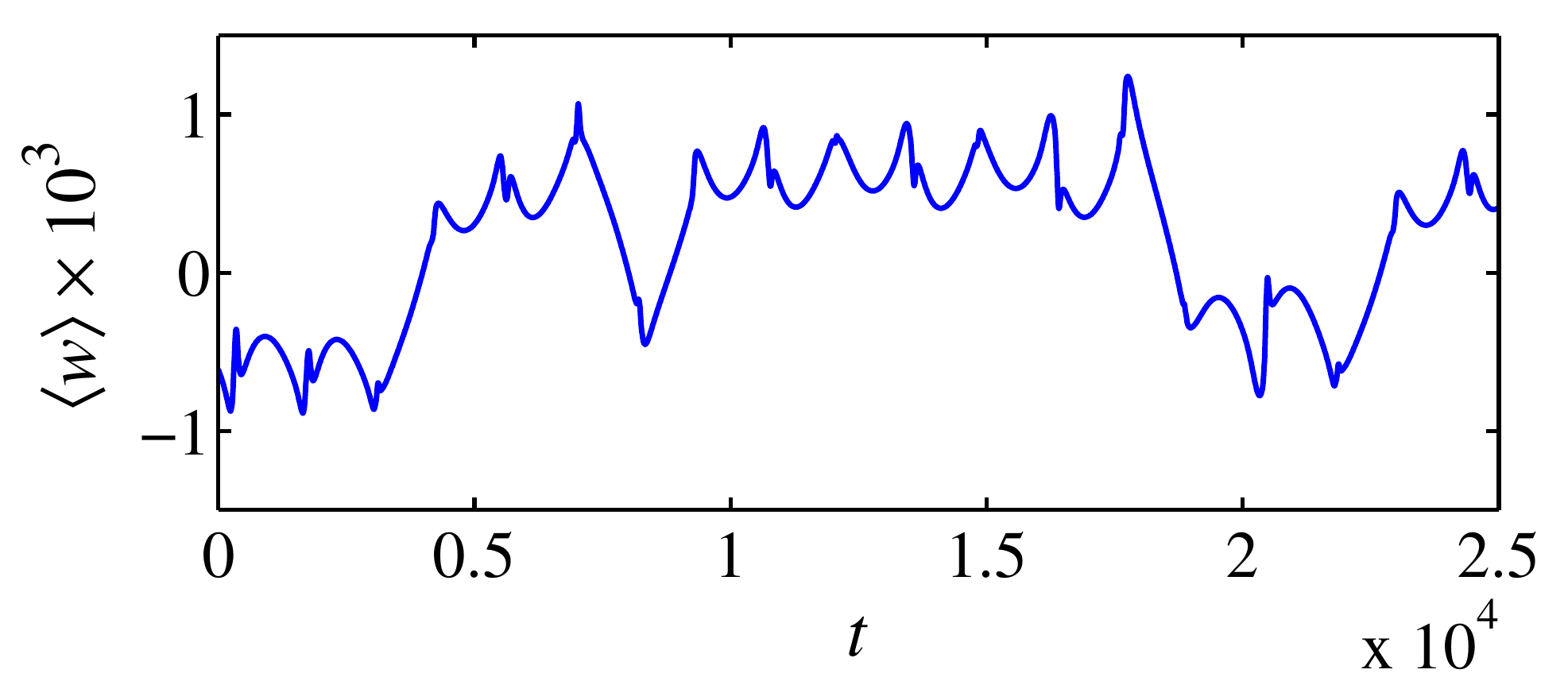}
\begin{picture}(0,0)
\put(-185,158){(\textit{a})}
\put(-185,75){(\textit{b})}
\end{picture}
\caption{\label{fig:4.5pi_Ecf} \small Time series for the erratic state at $L_x=4.5\pi$: (\textit{a})~cross-flow energy $E_{\mathrm{cf}}$;
(\textit{b}) mean
spanwise velocity $\langle w \rangle$.}
\end{figure}

Reducing $L_x$ further, no periodic states are identified on the L-branch between $L_x=4.6\pi$ and $L_x=4.45\pi$. The time signal of the
cross-flow energy, shown in fig.~\ref{fig:4.5pi_Ecf} for $L_x=4.5\pi$, is erratic, while still experiencing calm and bursting phases. As can be seen
from the space--time diagram in fig.~\ref{fig:long_st}(\textit{c}), the state does not change structurally, keeping the same characteristic
lengthscales and timescales. However its spatio-temporal dynamics is no longer regular but features a random walk, with unpredictable spanwise shifts
occurring at non-regular times. The various return maps did not reveal any clear structure and we refer to this state as "chaotic" or "erratic".
A phase-space representation of such a state is displayed in fig.~\ref{fig:phase-space}(\textit{b}).

\subsubsection{Period doubling ($L_x=4.3\pi$)}

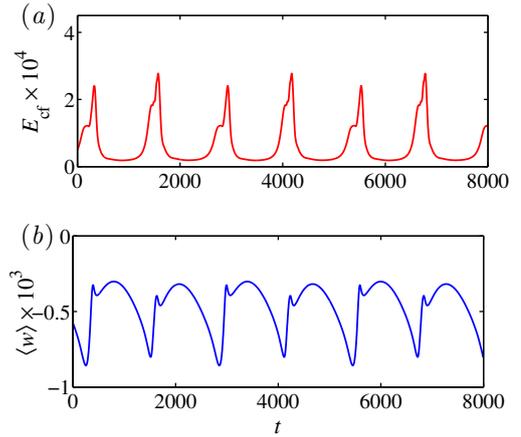
\begin{figure}
\centering
\includegraphics[scale=0.33]{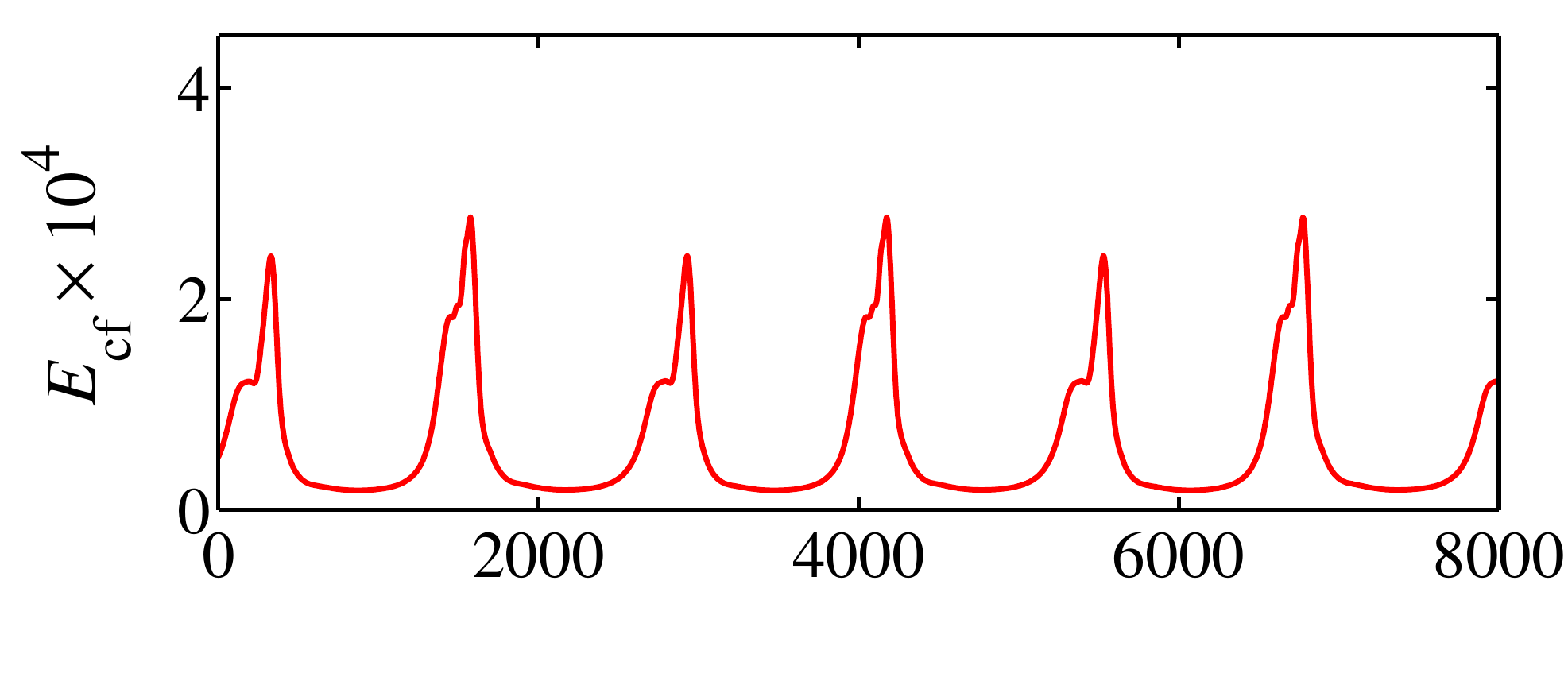}
\includegraphics[scale=0.33]{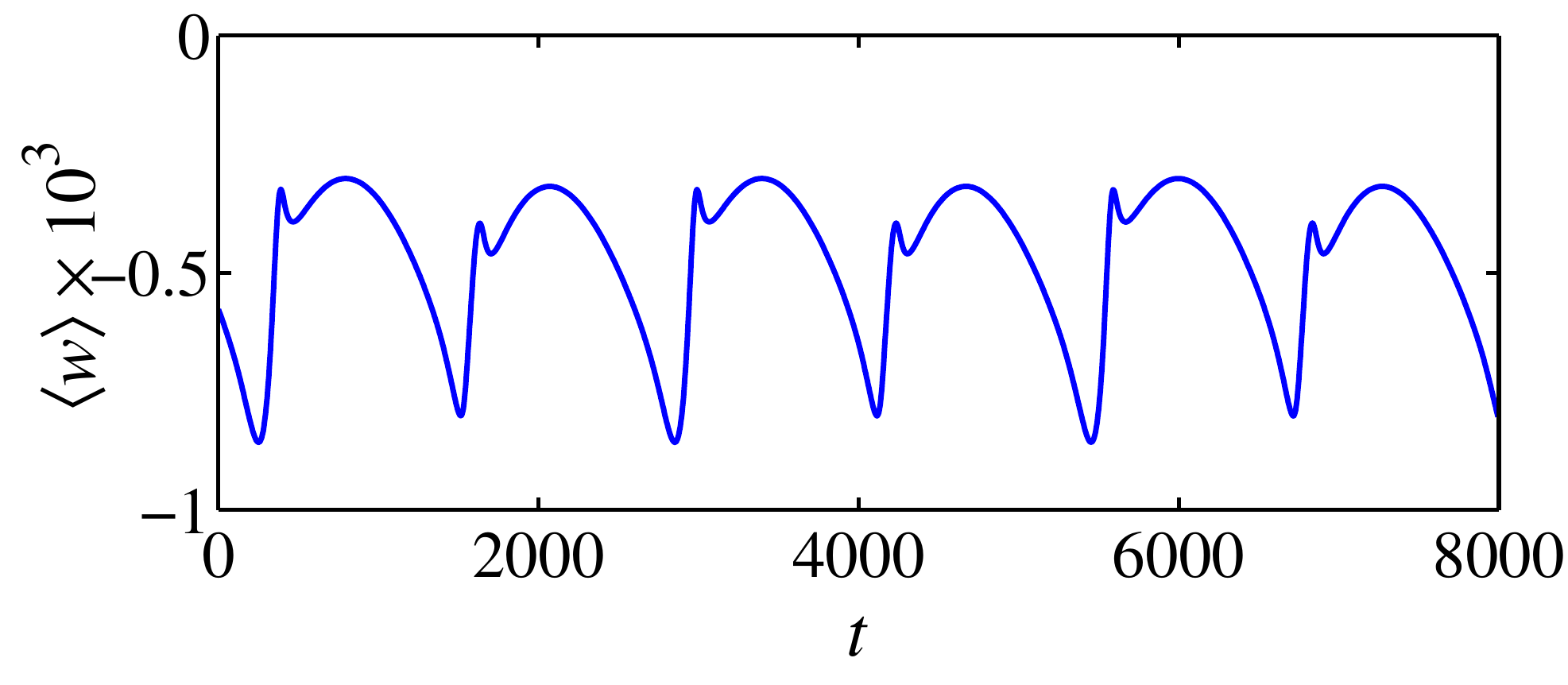}
\begin{picture}(0,0)
\put(-185,158){(\textit{a})}
\put(-185,75){(\textit{b})}
\end{picture}
\caption{\label{fig:4.3pi_Ecf} \small Time series for the $2$-periodic $\mathrm{L}^2$ state at $L_x=4.3\pi$: (\textit{a}) cross-flow energy
$E_{\mathrm{cf}}$; (\textit{b}) mean spanwise velocity $\langle w \rangle$.}
\end{figure}

Below $L_x=4.45\pi$ the state which repeatedly shifts only in one direction becomes attracting again. However unlike the previously identified L
state, two consecutive translations are no longer equivalent (see fig.~\ref{fig:4.3pi_Ecf}). The spatio-temporal dynamics consists of two shifts of
different lengths in the same direction. Their duration is different by less than $20\%$ and the total period of the state is approximately twice the
period found for larger $L_x$. We thus denote this edge state by $\mathrm{L}^2$ and call somewhat abusively "period-doubled". Somewhere between
$L_x=4.3\pi$ and $L_x=4.25$ the state undergoes a reverse period doubling bifurcation, and the original L state is recovered. It is stable down to
$L_x=4.2\pi$, below which the dynamics becomes chaotic again. Notable intermittent approaches to another periodic state were identified for
$L_x=4.15\pi$, as well as for $L_x=4.2\pi$ for other non-converging edge trajectories.

\subsection{Transition to turbulence from the edge states}

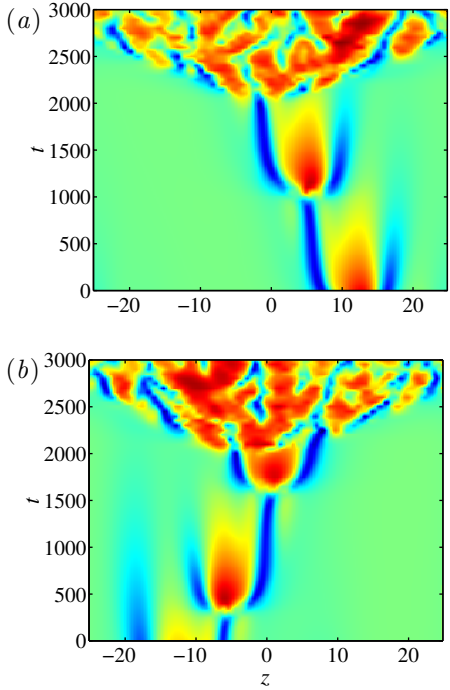
\begin{figure}
\centering
\includegraphics[scale=0.33]{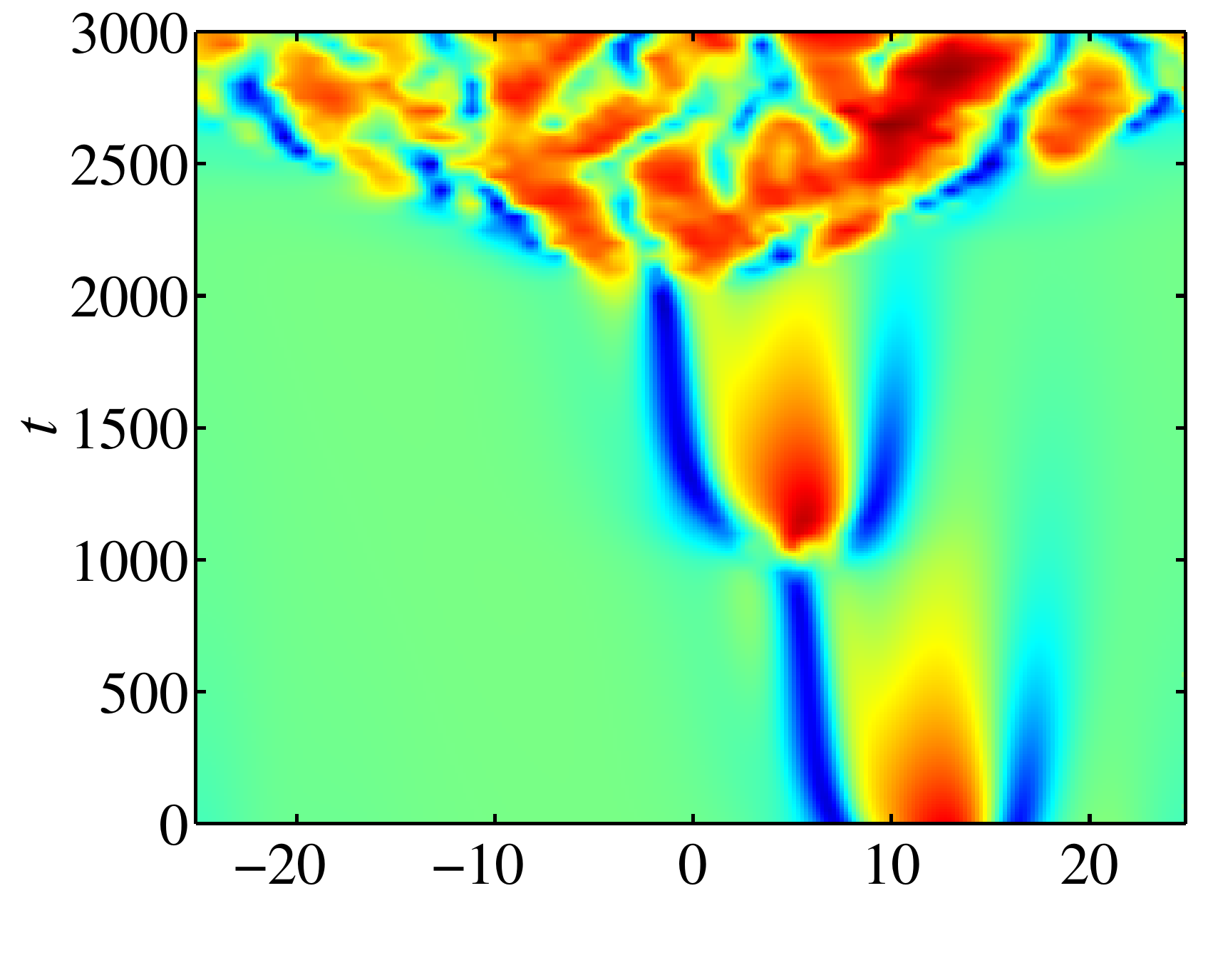}
\includegraphics[scale=0.33]{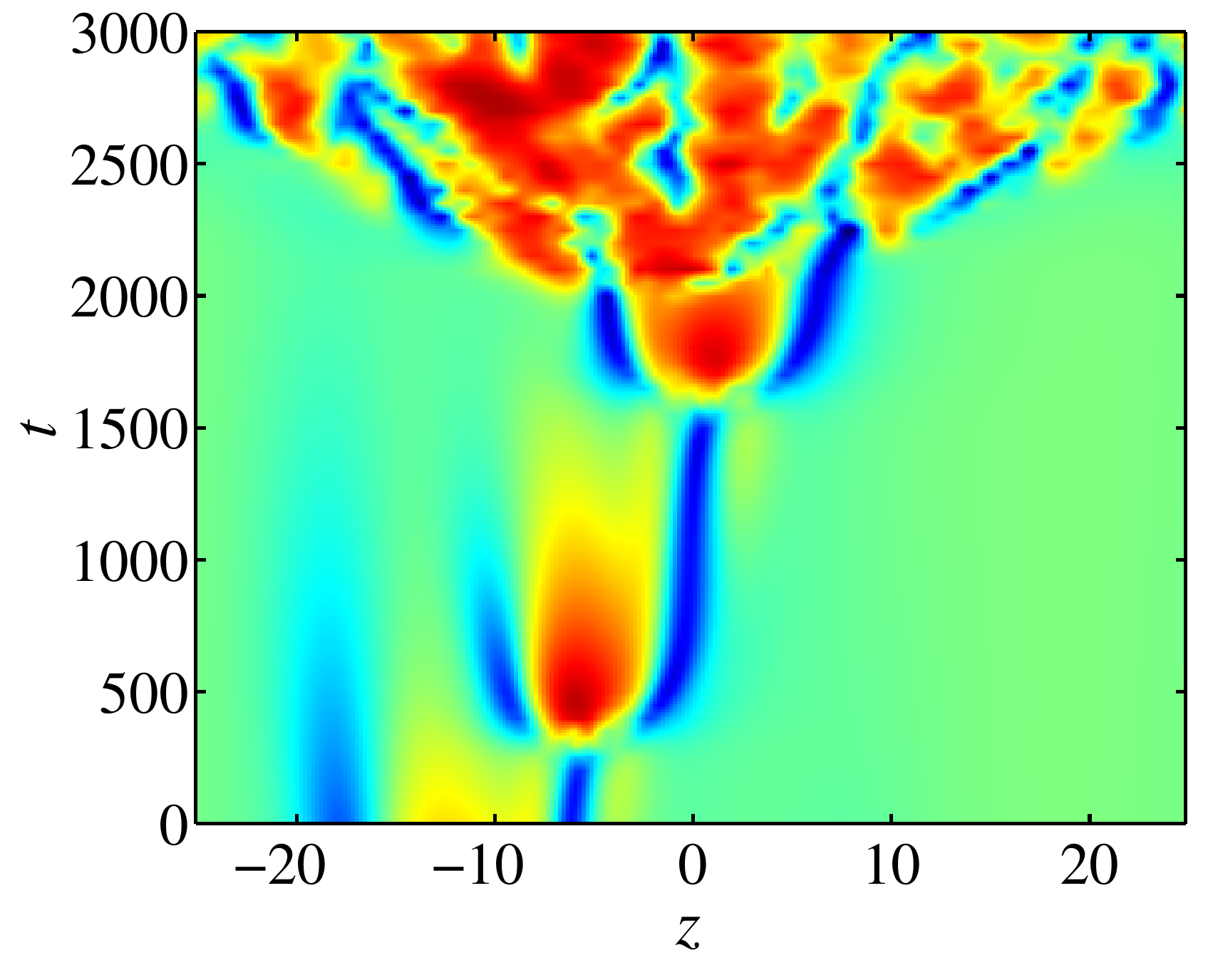}
\begin{picture}(0,0)
\put(-170,252){(\textit{a})}
\put(-170,120){(\textit{b})}
\end{picture}
\caption{\label{fig:turb_st} \small Space--time diagrams of the streamwise velocity fluctuations $u'$ averaged in $x$ at $y=1$ showing transition to
turbulence from (\textit{a}) L state at $L_x=6\pi$; (\textit{b}) Intermittent state at $L_x=4.7\pi$. The colourmap is the
same as in fig.~\ref{fig:6pi_st}.}
\end{figure}

\begin{figure}
\centering
\includegraphics[width=.9\linewidth]{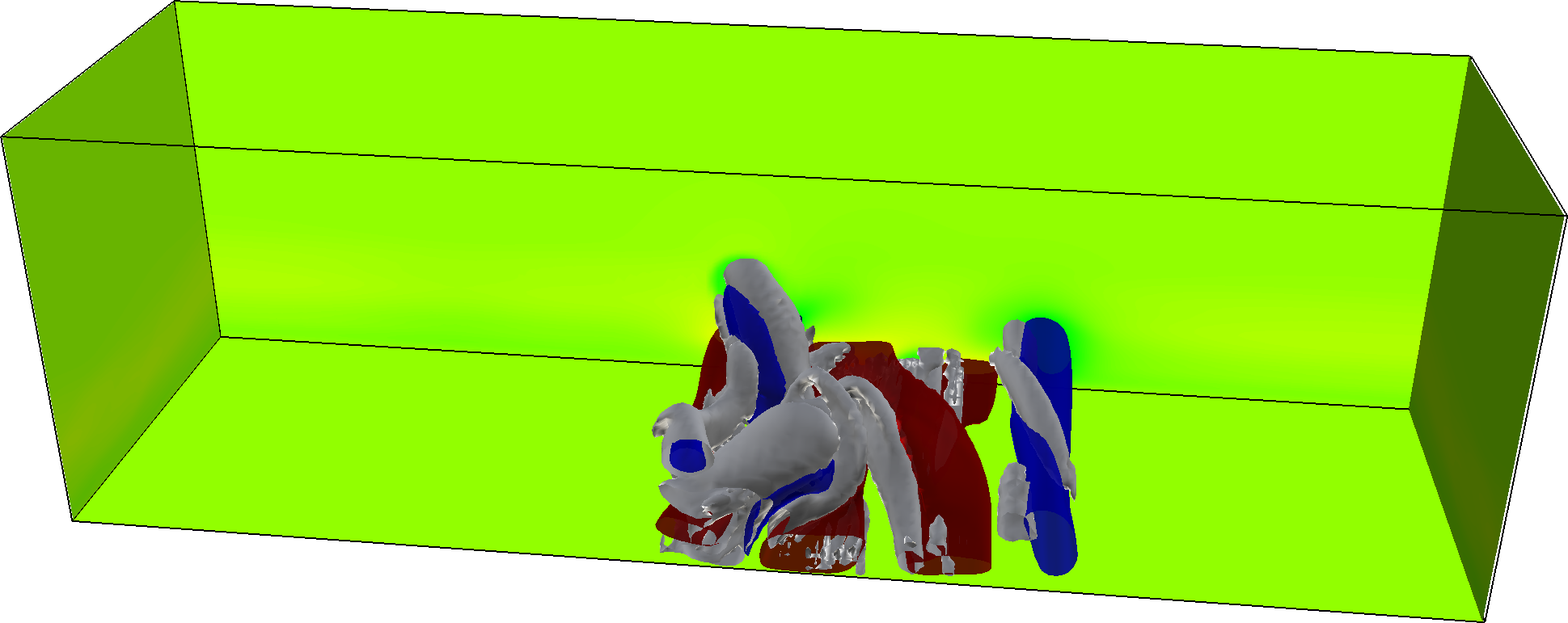}
\caption{\label{fig:edge_turb} \small Snapshot on the trajectory going turbulent of the $\mathrm{L}^2\mathrm{R}^2$ state for
$L_x=4.7\pi$
at $t=2,000$ in fig.~\ref{fig:turb_st}(\textit{b}). Isocontours are $u' = -0.2$ in blue, $u'=0.1$ in red and $\lambda_2=-0.001$ in grey.
The active left low-speed streak has already gone turbulent, whereas the right low-speed streak, which would decay when confined to the edge, also
develops sinuous instabilities. Flow from lower left to upper right.}
\end{figure}

Once the edge state is reached, simulations with initial conditions slightly above the separatrix produce trajectories which
stay for a finite time in the vicinity of the edge state and then rapidly leave towards the turbulent state along the leading unstable direction.
Since
the time scales of escape to turbulence are smaller then the scales of periodic motion that defines the states, the overall features of the transition
process are similar when starting from different states. Space--time diagrams depicting departure from edge state to turbulence are shown in
fig.~\ref{fig:turb_st}. Once the evolution leaves the edge state the structure quickly delocalises in $z$ and spreads over the full domain. This is
independent of where exactly the deviation from the edge trajectory occurs first. As seen in fig.~\ref{fig:edge_turb}, close to the edge state, not
only the active low-speed streak becomes turbulent, but also the one which decays when confined to the edge develops sinuous instabilities. In
contrast to Couette flow \cite{duguet_maitre_schlatter_2011} where the generated streaks have been found to be stationary in space, in the present
ASBL case the low- and high-speed streaks in the spreading turbulence are also drifting in the spanwise direction. Nucleation of new streaks appears
to be reminiscent of the spanwise translations of the discussed edge states.

\section{Discussion, perspectives}

When confined to the laminar--turbulent separatrix, the present system exhibits complex dynamics, with coexistence of more than one attractor.
The complexity raises questions about how those attracting states are connected together. They can either belong to different families of solutions or
be connected through local bifurcations. Multistability is typical in low-dimensional systems (see ref.~\cite{pisarchik_grebogi_2008} and related
references) and can be
also found in hydrodynamical settings (see \emph{e.g.}\ ref.~\cite{feudel_seehafer_tuckerman_gellert_2013}). For some
values of the parameter $L_x$ we identified here an unusual competition between a periodic and an erratic state. Performing edge state tracking
in this
situation, very long simulation times can be  needed to decide whether the
algorithm has converged to a chaotic state or is still converging to a simpler state. This demonstrates the limitations of the edge tracking procedure
for understanding and characterising the full structure of the laminar--turbulent boundary using a finite (small) number of simulations only.
Conversely, the linear stability properties of an edge state give no indication about the size of its basin of attraction, the latter being a fully
nonlinear
concept. In some situations this can lead to a misinterpretation of the dynamical role of some stable solutions, as pointed out in ref.~\cite{lebovitz_2009}
using a counterexample from a low-dimensional toy model.\\

Still, we were able to obtain a finer characterisation of the edge structure than, for instance, in ref.~\cite{duguet_willis_kerswell_2008}. We
suggest here a qualitative explanation of some exotic behaviours expected on the edge, such as multistability and intermittency. Our suggestion is
based on compiling results on exact coherent structures from other subcritical shear flows, notably the influence of varying either the streamwise or
spanwise wavelength of the numerical domain. The use of arc-length continuation robustly showed that most fundamental solutions such as travelling
waves or steady states are born and destroyed in saddle--node bifurcations
\cite{faisst_eckhardt_2003,wedin_kerswell_2004,gibson_halcrow_cvitanovic_2009}. Among these solutions, those with only
one unstable eigendirection
should in principle correspond to edge states. The coexistence of several loops of solutions in parameter space is typical and thus multistability
should be found on the edge when several solutions with only one unstable multiplier overlap in a given range of parameters. Such a situation has
already been reported in ref.~\cite{duguet_willis_kerswell_2008} for the travelling waves in pipe flow with $m=2$ symmetry. In the vicinity of these
saddle--node bifurcations, type-I intermittency \cite{pomeau_manneville_1980} should be found. In the present study we deal with
spanwise-localised solutions, hence the width of the domain $L_z$ is irrelevant once large enough. However we do report multistability in
some windows of the parameter $L_x$ and intermittency for other values of $L_x$.

\begin{figure}
\centering
\includegraphics[scale=0.4]{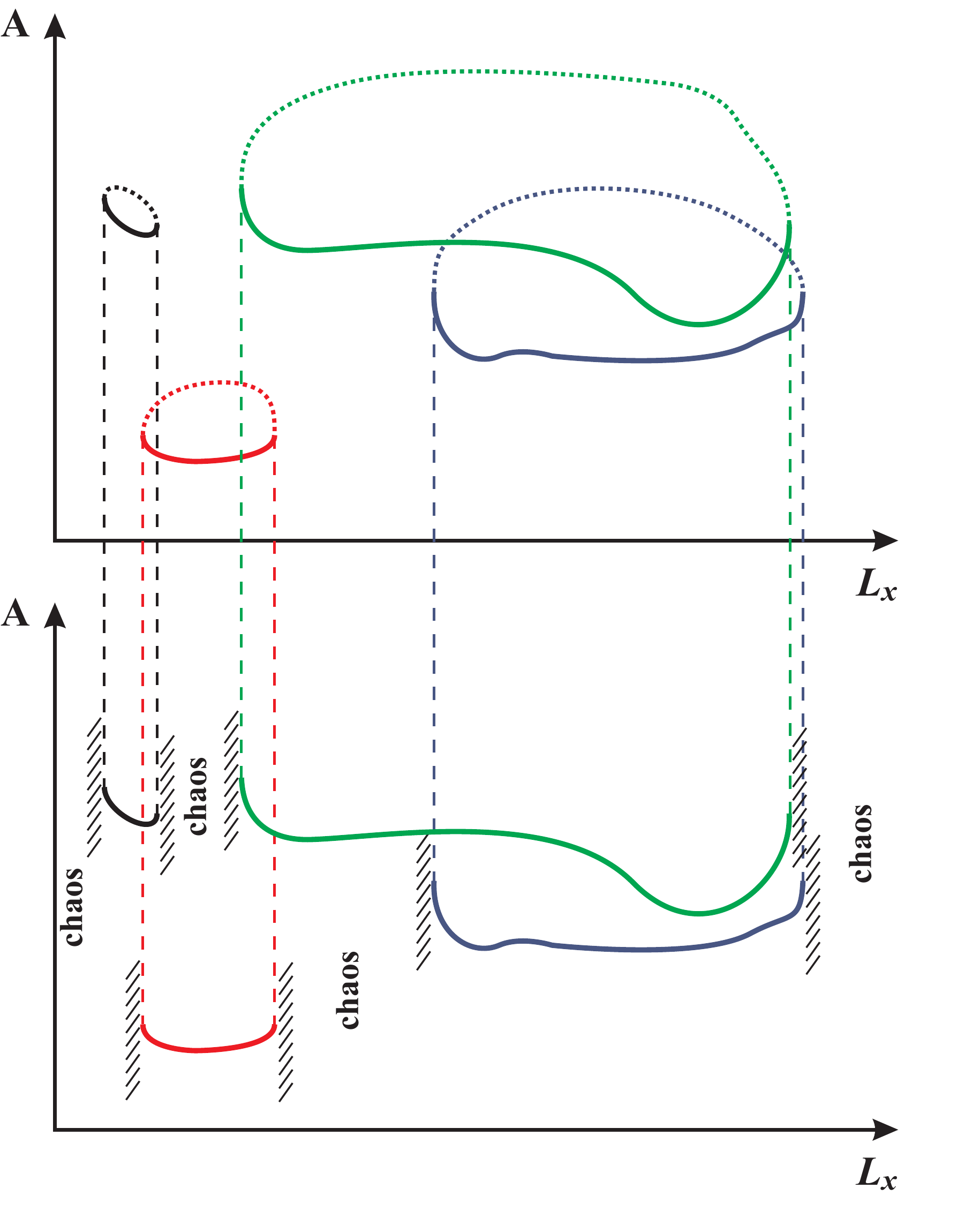}
\begin{picture}(0,0)
\put(-230,258){(\textit{a})}
\put(-230,131){(\textit{b})}
\end{picture}
\caption{\label{fig:sol_cont} \small Conceptual sketch of (\textit{a}) exactly periodic solutions as a function of $L_x$ and (\textit{b}) the
corresponding
bifurcation diagram for the dynamics on the edge. Solid lines represent solutions with one unstable direction, whereas dashed lines -- with more than
one. The thin vertical dashed lines connect the saddle--node bifurcation points between the upper and the lower part of the figure. Hatched areas
indicate possible intermittency zones.}
\end{figure}
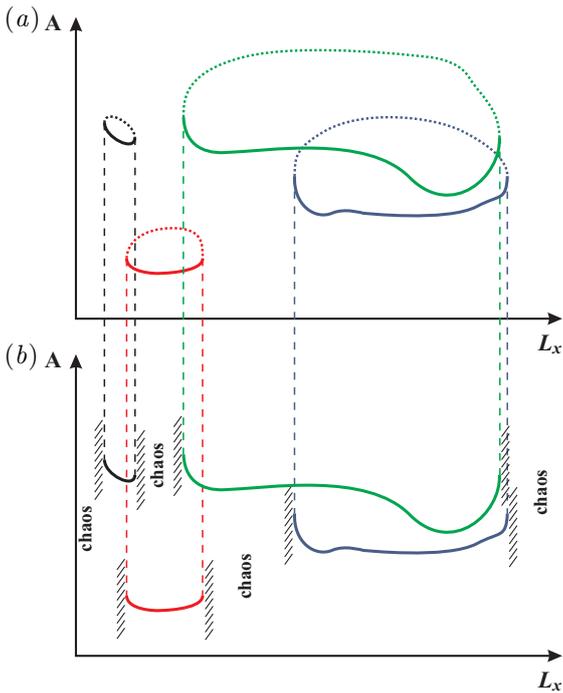

The above scenario is summarised graphically in fig.~\ref{fig:sol_cont}. We suppose (by direct analogy with former studies) that periodic solutions such
as those identified for $L_x=6\pi$ emerge as loops in fig.~\ref{fig:sol_cont}(\textit{a}). The $y$-axis refers to a suitable scalar observable, for
instance $E_{\mathrm{cf\_max}}$ (used in fig.~\ref{fig:bif_Lx}(\textit{a})) or $v_{\mathrm{rms}}$, the quantity used to iteratively find edge states.
Knowledge
of the stability properties of each branch (including possible local bifurcations along these branches) allows to list the solutions potentially behaving
locally as relative attractors on the edge. Locating the saddle--node bifurcations associated precisely with these solutions also allows for
predicting regions in $L_x$ where intermittency can be expected on the edge. This is shown in fig.~\ref{fig:sol_cont}(\textit{b}), which is
graphically deduced from fig.~\ref{fig:sol_cont}(\textit{a}) and compares conceptually well with the numerical
bifurcation diagram of fig.~\ref{fig:bif_Lx}(\textit{a}).

It is noteworthy that multistability and intermittency are not observed if the width of the domain $L_z$ is small. In
ref.~\cite{kreilos_veble_schneider_eckhardt_2013} edge states which are periodic in both wall-parallel directions are studied and the range in $L_x$
that we investigated here is also covered. However, only one edge state is found; it is only the spanwise localisation that allows multiple states to
coexist in the same domain.

We note, as many authors before us, that the total number of unstable directions of solutions lying on the edge is necessarily low, being equal to $1$
for edge states and finite otherwise. For the parameter values where edge trajectories stay erratic, consisting of transient approaches to
finite-amplitude solutions with at least $2$ unstable eigenvalues, the number of unstable Lyapunov exponents must also remain small. Moreover, the
robust spatial localisation of these edge states, verified in ref.~\cite{khapko_kreilos_schlatter_duguet_eckhardt_henningson_2013}, implies that
increasing the domain size in the direction of localisation does not modify the dynamics, and hence does not change the number of unstable Lyapunov
exponents even in chaotic regimes. The dynamics on the edge, though genuinely spatio--temporal, can hence safely be interpreted as an example of
non-extensive and low-dimensional dynamics. This has to be contrasted with the associated turbulent regime existing for the same parameters.  The
turbulent regime is not spatially localised and thus represents extensive chaos. Quantitative estimation of the number of unstable Lyapunov exponents
of typical turbulent trajectories is a hard task. A numerical estimate of this number was given in ref.~\cite{keefe_moin_kim_1992} in the case of a
``minimal" turbulent channel flow at $Re_{\tau}=180$, and was found to be $352$, \emph{i.e.}\ very large. Though analogies between all these different
flows are to be taken cautiously, the different orders of magnitude are unambiguous. The dynamics on the edge features a very small number of unstable
dimensions while the dynamics in the turbulent regime is highly unstable. This strongly supports the investigation of edge states as a laboratory for
investigating the dynamics and the self-sustenance mechanisms of coherent structures arising in turbulent flows.

We have here identified and characterised several edge states in spanwise-extended ASBL for different streamwise wavelengths. All these
states are structurally the same, consisting of an active localised pair of low- and high-speed streaks with streamwise vortices. The
dynamics of the streaks is built on the same elements as in the simplest L/LR cycles, \emph{i.e.}\ bursts and spanwise shifts, regardless of the
nature of the regime. Very similar coherent structures and dynamics were also found to be the minimal self-sustaining elements of near-wall
turbulence
\cite{jimenez_moin_1991,hamilton_kim_waleffe_1995,schoppa_hussain_2002,toh_itano_2003,jimenez_kawahara_simens_Nagata_shiba_2005}. Moreover, comparing
the three-dimensional visualisations of
both turbulence and the edge state in figs.~\ref{fig:turbulent_asbl} and~\ref{fig:6pi_structure}, similarities are obvious, let alone the localisation
properties, and closer comparison deserves more intensive investigation.

While edge states were first identified in small periodic domains, they have localised counterparts when the domain is extended in at least one
planar direction. The dynamics of such solutions usually mimics the one found in smaller domains (\emph{e.g.}\ bursts in cross-flow energy in the case
of ASBL). These localised solutions emerge from the spatially periodic branches either through a snaking scenario \cite{schneider_gibson_burke_2010}
or through large wavelength instabilities \cite{deguchi_hall_walton_2013,melnikov_kreilos_eckhardt_2013,chantry_willis_kerswell_2013}. The study in
ref.~\cite{chantry_willis_kerswell_2013} showed that streamwise-localised relative periodic orbits bifurcate off streamwise-periodic travelling waves
in pipe flow. By analogy we can thus expect that new fully localised solutions in ASBL bifurcate off the time-periodic spanwise-localised solutions
presented here.

In summary we have studied the dynamics on the lam\-inar--turbulent separatrix for a boundary-layer flow in a spanwise-extended set-up with
varying
streamwise periodicity. We were able to find a multitude of periodic and erratic edge states, which all share the same structure. The bifurcation
diagram obtained by compiling all these results contains a variety of interesting phenomena whose study is usually restricted to the framework of
low-dimen\-sional ODEs or iterated maps. Those include multistability, intermittency and period doubling. Besides their
properties as examples of simple dynamics embedded in a high-dimensional space, the identified states also constitute an interesting prototype flow
for understanding the self-sustaining mechanism of near-wall turbulence and developing nonlinear control strategies.

\section*{Acknowledgements}
T. Kh. would like to thank Paul Manneville and Predrag Cvitanovi\'c for discussions about the bifurcation diagram. Computer time
provided by SNIC (Swedish
National Infrastructure for Computing) is gratefully acknowledged.

\bibliographystyle{epj}
\bibliography{paperlist}

\end{document}

%% file: symbollines.tex
\font\smallfont=cmsy10 at 10truept
\mathchardef\bigCircle="280D

\font\bigfont=cmsy10 at 14.4truept
\mathchardef\tiMes="2902        %

\font\Bigfont=cmsy10 at 17.28truept
\mathchardef\DiaMond="2A05        %
\mathchardef\cirCle="2A0E
\mathchardef\BigCircle="2A0D

\font\Bbigfont=cmsy10 at 24.88truept
\mathchardef\buLLet="2B0F

\def\bigCirc{\raise 0.3ex\hbox{$\bigCircle$}\nobreak$\,$}

\def\Bullet{\raise-0.35ex\hbox{$\buLLet$}\nobreak$\,$}

\def\triangledown{\raise 0.2em\hbox{$\bigtriangledown$}\nobreak$\,$}

\def\minisquare{\hbox{${\vcenter{
               \hrule height 0.3pt \kern-0.4pt
               \hbox{\vrule width  0.3pt height 3.0pt \kern 2.6pt
               \vrule width  0.3pt height 3.0pt} \kern-0.4pt
               \hrule height 0.3pt}}$}}
\def\ssquare{\raise 0.175ex\hbox{${\vcenter{
               \hrule height 0.5truept       \kern-0.25truept
               \hbox{\vrule width 0.5truept height 3.0truept \kern 2.75truept
                     \vrule width 0.5truept height 3.0truept} \kern-0.25truept
               \hrule height 0.5truept}}$}\nobreak$\,$}
\def\squarex{\raise 0.175ex\hbox{${\vcenter{
               \hrule height 0.8truept       \kern-1.80truept
          \hbox{\vrule width 0.8truept height 8.0truept \kern-1.95truept
                \raise 0.8truept\hbox{$\tiMes$}     \kern-6.70truept
                \vrule width 0.8truept height 8.0truept} \kern-0.80truept
               \hrule height 0.8truept}}$}\nobreak$\,$}

\def\sqbull{\raise0.175ex\hbox{\vrule height 1.4ex width 1.6ex depth 0.2ex}\nobreak$\,$}
\def\smsqbull{\raise0.175ex\hbox{\vrule height 0.8ex width 0.9ex depth 0.2ex}\nobreak$\,$}

\def\Diamondplus{${\vcenter{\vcenter{\DiaMond} \kern-10truept
                            \hbox{\vrule width .4truept}\kern -3truept
                            \hrule height .4truept}}$\nobreak$\,$}
\newcount\ndots

\def\drawline#1#2{\raise 2.5truept\vbox{\hrule width #1truept height #2truept}}
\def\moonspace#1{\hskip #1truept}

\def\shortchain{\drawline{6.0}{0.75}}

\def\shortchainspace{\shortchain\moonspace{2}}

\def\Dashy{\drawline{4.00}{1.00}}     
\def\dashy{\drawline{4.00}{0.75}}     
\def\thindashy{\drawline{4.00}{0.25}}     
\def\dashyspace{\dashy\moonspace{2}}
\def\Dashyspace{\Dashy\moonspace{2}}
\def\thindashyspace{\thindashy\moonspace{2}}
\def\longdashy{\drawline{8.00}{0.75}} 
\def\thinlongdashy{\drawline{8.00}{0.25}} 
\def\longdashyspace{\longdashy\moonspace{2}}
\def\thinlongdashyspace{\thinlongdashy\moonspace{2}}
     
\def\dotty{\drawline{1.00}{0.75}}

\def\dottyspace{\dotty\moonspace{2}}

\def\solid{\drawline{24}{0.75}\nobreak$\,$}

\def\dashbox{\hbox{\dashyspace}}  
\def\Dashbox{\hbox{\Dashyspace}}  
\def\dashed{\hbox {\ndots=0 \loop\ifnum\ndots<3\advance\ndots by 1
        \dashbox\repeat\dashy}\nobreak$\,$}       
\def\Dashed{\hbox {\ndots=0 \loop\ifnum\ndots<3\advance\ndots by 1
        \Dashbox\repeat\Dashy}\nobreak$\,$}       
\def\thindashbox{\hbox{\thindashyspace}}  
\def\thindashed{\hbox {\ndots=0 \loop\ifnum\ndots<3\advance\ndots by 1
        \thindashbox\repeat\thindashy}\nobreak$\,$}       
\def\thindash{\hbox {\ndots=0 \loop\ifnum\ndots<3\advance\ndots by 1
        \thindashbox\repeat\thindashy}\nobreak$\,$}       

\def\longdashbox{\hbox{\longdashyspace}}  
\def\thinlongdashbox{\hbox{\thinlongdashyspace}}  
\def\longdash{\hbox {\ndots=0 \loop\ifnum\ndots<3\advance\ndots by 1
        \longdashbox\repeat\longdashy}\nobreak$\,$}       
\def\thinlongdash{\hbox {\ndots=0 \loop\ifnum\ndots<3\advance\ndots by 1
        \thinlongdashbox\repeat\thinlongdashy}\nobreak$\,$}       

\def\dotdashed{\hbox{\shortchainspace\dottyspace\shortchain}\nobreak$\,$}      

